\documentclass{article}

\usepackage{arxiv}

\usepackage[utf8]{inputenc} 
\usepackage[T1]{fontenc}    
\usepackage{hyperref}       
\usepackage{url}            
\usepackage{booktabs}       
\usepackage{amsfonts}       
\usepackage{nicefrac}       
\usepackage{microtype}      
\usepackage{graphicx}
\usepackage{doi}

\usepackage{authblk}
\usepackage{amsmath, bm}
\usepackage{upgreek}
\usepackage{longtable}
\DeclareFontFamily{OT1}{pzc}{}
\DeclareFontShape{OT1}{pzc}{m}{it}{<-> s * [1.500] pzcmi7t}{}
\DeclareMathAlphabet{\mathpzc}{OT1}{pzc}{m}{it}
\usepackage[bibstyle=numeric,sorting=none,firstinits=true]{biblatex}
\graphicspath{ {./figure/} }

\title{A Physics-Guided Neural Network Framework for Elastic Plates: Comparison of Governing Equations-Based and Energy-Based Approaches}

\author{Wei Li\textsuperscript{ a} 
        \quad  Martin Z. Bazant\textsuperscript{ b, c}
        \quad Juner Zhu\textsuperscript{ a, b,} \thanks{Corresponding author.
Emails: weili17@mit.edu (W.L.), bazant@mit.edu (M.Z.B), zhujuner@mit.edu (J.Z.)}
        \\
        \textsuperscript{a}Department of Mechancial Engineering, Massachusetts Institute of Technology \\
        \textsuperscript{b}Department of Chemical Engineering, Massachusetts Institute of Technology\\
        \textsuperscript{c}Department of Mathematics, Massachusetts Institute of Technology}

\begin{document}
\maketitle

\begin{abstract}
	One of the obstacles hindering the scaling-up of the initial successes of machine learning in practical engineering applications is the dependence of the accuracy on the size of the database that ``drives'' the algorithms. Incorporating the already-known physical laws into the training process can significantly reduce the size of the required database. In this study, we establish a neural network-based computational framework to characterize the finite deformation of elastic plates, which in classic theories is described by the Föppl--von Kármán (FvK) equations with a set of boundary conditions (BCs). A neural network is constructed by taking the spatial coordinates as the input and the displacement field as the output to approximate the exact solution of the FvK equations. The physical information (PDEs, BCs, and potential energies) is then incorporated into the loss function, and a pseudo dataset is sampled without knowing the exact solution to finally train the neural network. The prediction accuracy of the modeling framework is carefully examined by applying it to four different loading cases: in-plane tension with non-uniformly distributed stretching forces, in-plane central-hole tension, out-of-plane deflection, and buckling under compression. Three ways of formulating the loss function are compared: 1) purely data-driven, 2) PDE-based, and 3) energy-based. Through the comparison with the finite element simulations, it is found that all the three approaches can characterize the elastic deformation of plates with a satisfactory accuracy if trained properly. Compared with incorporating the PDEs and BCs in the loss, using the total potential energy shows certain advantage in terms of the simplicity of hyperparameter tuning and the computational efficiency.
\end{abstract}


\section{Introduction}\label{introduction}

In the past half-decade, machine learning enjoyed vast researches to achieve remarkable successes in a wide spectrum of scientific problems, including image processing \cite{Egmont-Petersen2002Image,Rawat2017Deep}, cognitive science \cite{French2002Introduction}, genomics \cite{Libbrecht2015Machine}, drug discovery \cite{Lo2018Machine}, and material designing \cite{Ramprasad2017Machine}, to name a few. It has shown prominent advantages over other methods in effectively handling complex natural systems with a daunting number of variables. Recently, we are witnessing a growing number of initial successes of machine learning, especially deep learning, in modeling complex engineering systems with a high dimensionality (the number of variables and degrees of freedom) \cite{Alber2019Integrating, Han2018Solving}, for example, predicting the remaining useful life of a battery cell based on its partial life-cycle data \cite{Severson2019Data-driven}. In most cases, machine learning algorithms serve as a data-driven approach. It has been proven effective to predict the performance of a system even when the underlying physics has not been fully elucidated. However, like other statistical methods such as curve fitting and feature engineering, the accuracy of machine learning algorithms highly depends on the quantity and quality of the dataset used for training \cite{Famili1997Data}. In the cases \noindent where little data is accessible or a large database is not affordable, for example at the microscales and nanoscales of unknown materials, machine learning algorithms may lose their power, thus hindering the scaling-up of those initial successes. On the contrary, physics-based or first-principle-based models are commonly less reliant on the size of the dataset because the governing physical laws are elucidated by human brains beforehand and only a small amount of data is required to calibrate the unknown parameters.

Bridging the gap between the data-driven approaches and the physics-based approaches creates a promising opportunity to develop novel computer methods that have the potential to unite the advantages of both approaches -- characterizing high-dimensionality systems with a small dataset. One type of these methods is often referred to as physic-guided data-driven methods, which aim to implement the already-known physics into the data-driven approaches \cite{Han2018Solving,Sirignano2018DGM:,Wang2020mesh-free,Weinan2018Deep,Raissi2019Physics-informed,Wang2017Physics-informed,Samaniego2020energy,Karpatne2017Theory-guided,long2019pde}. In this paper, we focus on machine learning algorithms, particularly artificial neural networks (ANNs). Generally, there are three key elements in a machine learning application: a dataset, a model, and a training process. In the vast open literature, it is found that the term ``physics-guided machine learning'' (PGML) is being used in an unregulated manner. Here, we summarize the recent progress on this topic by classifying the existing studies in the open literature according to which key element they worked on.

The first category of studies implements physics into machine learning algorithms by generating a dataset following physical laws \cite{Li2019Data-Driven,Chen2020Direct}. Instead of collecting data from the expensive and time-consuming experiments, attempts were made to generate data through first-principle theories and physics-based simulations. In this way, the PGML algorithm can learn from the known physics behind the man-made data. For example, Chen et al. \cite{Chen2020Direct} predicted the phonon density of states of crystalline solids with unseen elements using a density functional perpetration theory-based phonon database to train a Euclidean neural network. Another example is that Li et al. \cite{Li2019Data-Driven} generated a large numerical dataset of lithium-ion battery failure behavior under mechanical impact loads with a well-calibrated physics-based detailed model and trained various machine learning algorithms to get the safety envelope of the cell.

The second category of PGML studies implements physical laws by designing a physics-guided machine learning model \cite{Han2018Solving,Zhang2018DeePCG:,Darbon2020Overcoming}. Various types of models have been used and studied for machine learning systems to solve a target regression or classification problem, for example, ANNs and support vector machines (SVMs). Many existing studies treated the model as a black box and empirically choose a set of parameters of the model, such as the number of nodes and hidden layers of ANNs. But it has become a consensus that understanding the physics of the problem can guide the design of the model. E's research team \cite{Han2018Solving,Zhang2018DeePCG:} is a clear pioneer in this aspect. In one of their recent studies \cite{Han2018Solving}, a neural network was designed with several subnetworks representing the solution at different time instances to solve the high-dimensional partial differential equations (PDEs). Several important physical equations were successfully solved with their algorithm, including the nonlinear Black--Scholes equation with default risk, the Hamilton--Jacobi--Bellman equation, and the Allen--Cahn Equation.

The last category of PGML studies imposes physics into the training process of the algorithm. A typical example is the ``physics-informed neural network'' (PINN) proposed by Raissi et al. \cite{Raissi2019Physics-informed} PINN introduces the PDEs and the associated initial and boundary conditions (ICs and BCs) into the loss function and solves the PDEs by minimizing it. The authors successfully applied the PINN approach to solving the Burger's equation and the Navier-Stokes equation, which are the governing equations of a variety of flows. The same idea was adopted by Lu et al. \cite{Lu2019DeepXDE:} to solve a series of PDEs such as the diffusion equation, and a library of open-source codes for solving different PDEs was created by the authors, named ``DeepXDE''. Besides PINN, E et al. \cite{Weinan2018Deep} proposed a deep learning Ritz method \noindent where the loss function is defined by the energy functional corresponding to the PDEs. It is also found that this type of loss-function-based optimization algorithms can not only  predict the performance of a system by solving the governing equations but also identify the unknown parameters in a physical law through an inverse process. Raissi et al \cite{Raissi2019Physics-informed} showed preliminary successes to discover the unknown parameters in the Burger's equation using PINN. In fact, this type of general applications of data-driven methods does not rely on machine learning algorithms. Zhao et al. \cite{Zhao2020Learning} used an inverse approach to learn pattern-forming equations such as Cahn-Hilliard and Allen-Cahn from image data. Zhao's approach turned out to be still effective even with a very small set of images. This success was recently extended by Effendy et al \cite{Effendy2020Analysis} to analyze and design the electrochemical impedance spectroscopy of energy storage systems.

Although we classify here the existing PGML studies in the open literature into three categories, it is worth noting that there is no absolute boundary between them. It is possible to combine these approaches in one algorithm, and with the rapid development of machine learning technologies, we are already witnessing an increasing number of advanced algorithms that have all the above three merits \cite{E2020Integrating,Qian2020Lift}.

There is a recent trend that the success of physics-guided machine learning algorithms, which was initially achieved in the fields like fluid dynamics and mass and heat transfer, is now being extended into the field of applied mechanics of solids. Haghighat et al. \cite{Haghighat2020deep} applied the PINN framework to predicting the mechanical responses of linear elastic materials, and their predictions agreed well with the finite element simulations. Wu et al. \cite{Wu2020recurrent} designed a recurrent neural network-accelerated multiscale model to describe the elasto-plastic behavior of heterogeneous media subjected to random cyclic and non-proportional loading paths. A recent study by Samaniego et al. \cite{Samaniego2020energy} adopted an energy approach (variational method) to solve the PDEs in solid mechanics with machine learning and shows a high prediction accuracy for the given examples. Huang et al. \cite{Huang2020machine} developed a machine learning-based plasticity model that can effectively predict the behaviors of history-dependent materials. In our opinion, one of the challenges for the machine learning applications in predicting the mechanical responses of solids is that most of the variables of interest (such as stress and strain) are highly tensorial, or multi-axial. As a result, each direction has its own PDEs, leading to a large total number of equations as well as BCs and ICs. Therefore, although the above studies all compared their predictions with other numerical methods and showed satisfactory agreements, the most fundamental questions such as how the loss function should be formulated are still not full settled.

The purpose of the present study is to develop a neural network framework for predicting the mechanical responses of elastic plates using the third category of the aforementioned PGML strategies (implementing physics into the training process). Our work will be distinguished from the existing publications in a number of ways. First, we will carefully investigate the reliability of the neural network framework by solving a set of high-order and highly non-linear equations. In the classic theories of elastic plates, the governing equations are the well-known Föppl-von Kármán equations equations, which consists of two second-order PDEs and one fourth-order PDE. Second, two different approaches will be used to construct the loss function, one based on PDEs and BCs (inspired by PINN), and the other based on the total potential energy of the whole structure (inspired by deep Ritz). Third, the proposed computational framework will be applied to four different loading cases of the plates for evaluation. Non-linearities that stem from the loading condition and the geometry of the plates will be purposely introduced to push the developed numerical framework to its limit. The paper will start with a brief introduction of the classic Kirchhoff plate theory. The theory of the neural network framework is then presented, together with a comparison with the conventional purely date-driven machine learning framework. At last, four exemplary loading conditions will be investigated, and some key features of the framework will be discussed.

\section{Physics to be implemented: theory of elastic plates}\label{theory-of-elastic-plates}

The governing equations of an elastic plate can be obtained in two ways. One is using the condition for adjacent equilibrium, and the other is following the energy method. The former applies Newton's laws on a control volume of the structure by balancing all the applied and reactive forces and moments in each direction. This is quite straightforward for simple mechanical systems \noindent where the forces and displacements are easy to be described. For complicated systems and structures, however, using the energy method is usually more convenient. It calculates the total potential energy as a functional and establishes the equilibrium equations (Euler-Lagrange equations of the functional) based on its variations. From a theoretical point of view, these two ways of obtaining the governing equations are equivalent up to regularity considerations. Here, the energy method will be used.

    \subsection{Principle of virtual displacements}\label{principle-of-virtual-displacements}

    For a given mechanical system, there are infinite possible configurations that can satisfy the geometric constraints. However, only the one that also satisfies the equilibrium condition is the true configuration. The displacement field corresponding to the true configuration is the true displacement, and the virtual displacements represent all the possible configurations consistent with the geometric constraints. The amounts of the virtual work is then the work done by all the forces along with the virtual displacements. The virtual work done by the internal stress and external body or surface force are respectively defined as internal virtual work \(\updelta U\) and external virtual work \(\updelta V\). Among all the admissible configurations, the one corresponding to the equilibrium configuration makes the total virtual work \(\updelta \Pi\) vanish. The principle of virtual displacement is then stated as

	\begin{equation} 
		\label{eq:eq1}
		\updelta \Pi = \updelta U + \updelta V = 0,
	\end{equation}

    \noindent \noindent where \(\updelta\) is the variation operator. From this equation, we will derive the governing equations of the plates in the following by writing down the internal and external work and making use of the principles of the variation method.

    \subsection{Displacement field and strains}\label{displacement-field-and-strains}

    We adopt the classic plate theory that is based on the following three Kirchhoff hypotheses \cite{Reddy2006}: i) straight lines normal to the plate mid-surface remain normal after deformation and thus are also called transverse normals; ii) the transverse normals remain straight after defter deformation, and iii) the thickness of the plate does not change after deformation.

    \begin{figure}
		\centering
		\includegraphics[width=0.8\columnwidth]{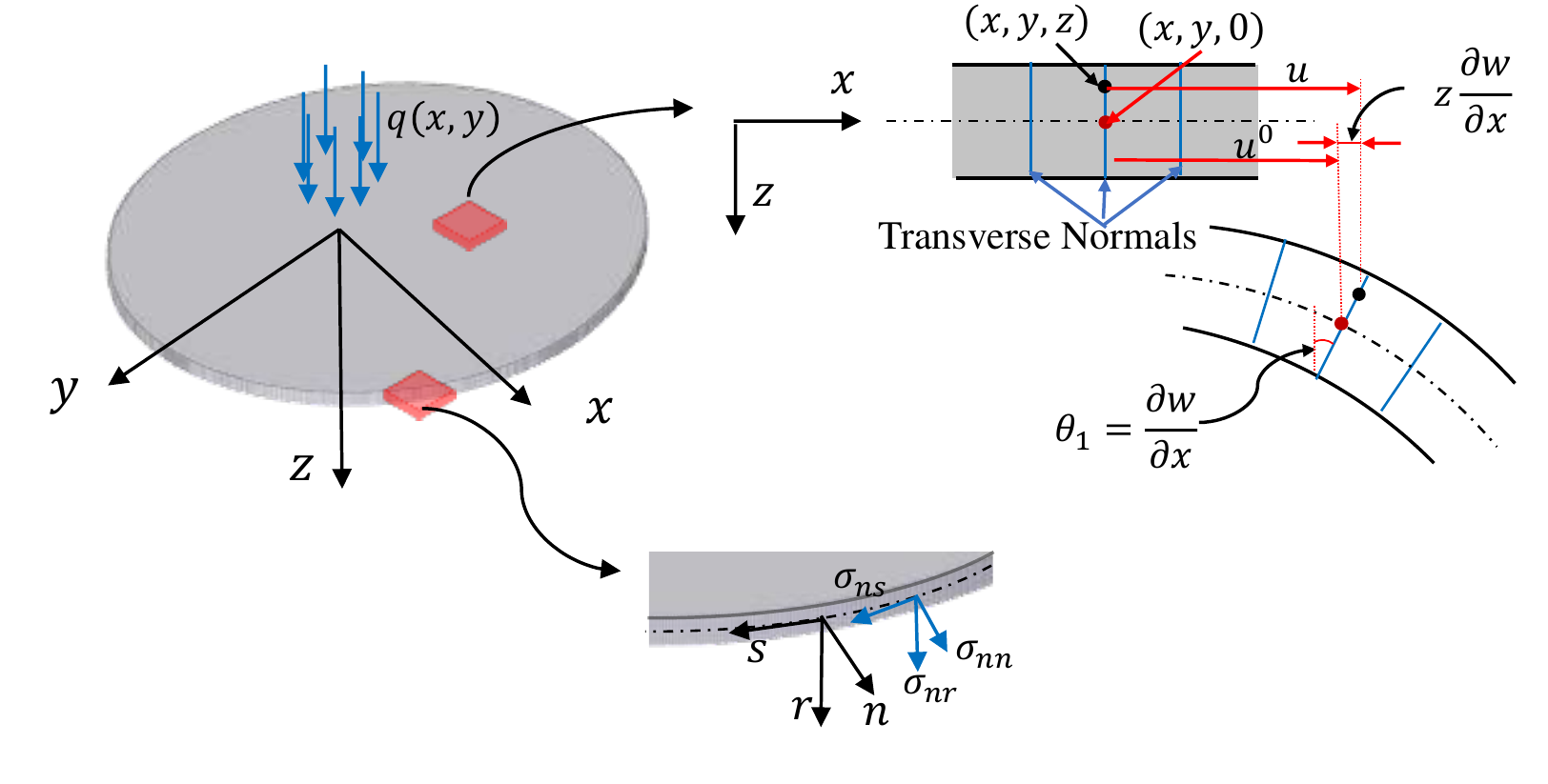}
		\caption{Illustration of the reference and deformed configurations of an elastic plate and the
		boundary conditions}
		\label{fig:fig1}
	\end{figure}

    Figure~\ref{fig:fig1} shows a plate of thickness \(h\) in the Cartesian coordinate \((x,y,z)\). The \(x\)-\(y\) plane coincides with the geometric mid-plane of the plate and the \(z\)-direction is taken positive downward. Without loss of generality, the three components of displacement field along the \(x\), \(y\), and \(z\) axes are noted as \(u\), \(u_y\), and \(w\), respectively. Based on the Kirchhoff hypotheses, the strain components can be written as (refer to \ref{Appendix A.1} for a detailed derivation),

	\begin{equation} 
		\label{eq:eq2}
			\varepsilon_{\alpha\beta} = \varepsilon_{\alpha\beta}^{0} + z\cdot\kappa_{\alpha\beta},
	\end{equation}

    \noindent where \(\varepsilon_{\alpha\beta}^{0}\) \((\alpha, \beta = 1, 2)\) are the membrane strains that represent the in-plane deformation, and \(\kappa_{\alpha\beta}\) \((\alpha, \beta = 1, 2)\) are the curvatures (often known as the bending strains) that comes from the transverse bending,

	\begin{equation} 
		\label{eq:eq3}
            \varepsilon_{\alpha\beta}^0 = \frac{1}{2}(u^0_{\alpha,\beta} + u^0_{\beta, \alpha} + w_{,\alpha}w_{,\beta}),
	\end{equation}
	
	\begin{equation} 
		\label{eq:eq3-2}
            \kappa_{\alpha\beta} = - w_{,\alpha\beta},
	\end{equation}

    \noindent where the comma notation ``,''  indicates a derivative (for example, \(w_{,\alpha} = {\partial w}/{\partial \alpha} \)), \(u^0_\alpha\) \((\alpha=x,y)\) are the displacements on the mid-plane (\(u^0_{\alpha}(x_{\alpha}) = u_\alpha(x_{\alpha},0)\)). \(w_{,\alpha}\) \((\alpha=x,y)\) are respectively the rotation angles of the transverse normal along \(x\) and \(y\) axes.

    \subsection{Potential energy and variation method}\label{potential-energy-and-variation-method}

    The internal virtual work (variation of internal energy) of a plate is defined as
    
    \begin{equation}
		\label{eq:eq4}
		\updelta U = \int_{\Omega}^{}{
			\left( N_{\alpha\beta}\updelta\varepsilon_{\alpha\beta}^0 + M_{\alpha\beta}\updelta\kappa_{\alpha\beta} \right) \text{d}x\text{d}y},
	\end{equation}
    \noindent where \(N_{\alpha\beta}\)  and \(M_{\alpha\beta}\) are respectively the thickness-integrated forces and moments (also known as the membrane forces and bending moments),

    \begin{equation} 
        \label{eq:eq5}
			N_{\alpha\beta} = \int_{-\frac{h}{2}}^{\frac{h}{2}}{\sigma_{\alpha\beta} \text{d}z},
    \end{equation}
    
    \begin{equation}
        \label{eq:eq6}
        M_{\alpha\beta} = \int_{-\frac{h}{2}}^{\frac{h}{2}}{\sigma_{\alpha\beta}z \text{d}z}.
    \end{equation}

    We consider a distributed transverse (\(z\)-direction) pressure \(q_{t}\) on the top surface and a set of mixed traction-displacement boundary conditions. For generality, we define a local coordinate \((n,\ s,\ z)\) at any point on the boundary where \(n\) and \(s\) are corresponding to the normal and tangential directions. The applied in-plane traction can thereby be described as the normal stress \({\hat{\sigma}}_{\text{nn}}\), tangential stress \({\hat{\sigma}}_{\text{ns}}\), and the transverse shear stress \({\hat{\sigma}}_{\text{nz}}\). Hence, the external virtual work can be calculated as

    \begin{equation}
		\label{eq:eq7}
		\begin{aligned}
			V
		      &= - \int_{\Omega}^{}{q_{t}\updelta w}\text{d}x\text{d}y - \int_{\Gamma_{\sigma}}^{}{\left( {\widehat{N}}_{nn}\updelta u_{0n} - {\widehat{M}}_{nn} \updelta w_{,n} + {\widehat{N}}_{ns}\updelta u_{0s} - {\widehat{M}}_{ns} \updelta w_{,s} + {\widehat{N}}_{nz}\updelta w \right) \text{d}s},
		\end{aligned}
	\end{equation}

    \noindent where \(u_{n}\) and \(u_{s}\) are respectively the displacements along the boundary normal and tangential direction, \(u_{0n}\) and \(u_{0s}\) are respectively the corresponding displacements at the mid-plane, \(w_{,n}\) and \(w_{,s}\) are respectively the rotation angels of the transverse normal along the boundary normal and tangential directions, and the applied thickness-integrated forces and moments are defined in the same way as in Eq. \eqref{eq:eq5} and Eq. \eqref{eq:eq6}. 

    To express the variation of internal energy in terms of virtual displacements, integration by parts need to be performed several times (details can be found in \ref{Appendix A.2}). According to the principle of virtual displacements and rearranging the coefficients, we finally have

    \begin{equation}
		\label{eq:eq8}
		\begin{aligned}
		0 & = \updelta U + \updelta V \\
		  &  = - \int_{\Omega}^{} \left\{ N_{\alpha\beta,\beta}\updelta u_{\alpha}^0 + \left[\left(N_{\alpha\beta}w_{,\beta}\right)_{,\alpha}
			 + M_{\alpha\beta,\alpha\beta} - q_{t} \right] \updelta w  \right\} \text{d}x\text{d}y \\
		    & + \int_{\Gamma_{\sigma}}^{}\Bigl[ N_{\alpha\beta}n_{\beta} \updelta u_{\alpha}^0
			 + \left(N_{\alpha\beta}w_{,\beta}n_{\alpha}
			 + M_{\alpha\beta,\beta}n_{\alpha} \right)\updelta w
			- M_{\alpha\beta}n_{\beta}\updelta w_{,\alpha} \\
			& + {\widehat{N}}_{\text{nn}}\updelta u_{0n} - {\widehat{M}}_{\text{nn}}\frac{\partial\updelta w}{\partial n} 
			 + {\widehat{N}}_{ns}\updelta u_{0s} - {\widehat{M}}_{ns}\frac{\partial\updelta w}{\partial s} + {\widehat{N}}_{nz}\updelta w \Bigr] \text{d}s.
		\end{aligned}
	\end{equation}

    The right-hand-side of Eq. \eqref{eq:eq8} consists of two integration terms. The first is over the whole domain \(\Omega\), and the second is along the traction-based BCs \(\Gamma_{\sigma}\). The principle of virtual displacement, \(\updelta \Pi = 0\), implies that any small changes in the displacement field should not change the totally potential energy. Mathematically, both integration terms should be zero. Considering that the first term is zero, the coefficients of \(\updelta u_{\alpha}^0\) (\(\alpha=x,y\)), and \(\updelta w\) must all be zero. We can get the following governing equations (i.e. the Euler-Lagrange equations of the total potential energy functional),

    \begin{equation}
		\label{eq:eq9}
		\begin{aligned}
			\mathcal{P}_{\alpha}& \equiv N_{\alpha\beta,\beta} = 0, \\
			\mathcal{P}_z& \equiv \left(N_{\alpha\beta}w_{,\beta}\right)_{,\alpha}
			 + M_{\alpha\beta,\alpha\beta} - q_{t} = 0.
		\end{aligned}
	\end{equation}

    Eq. \eqref{eq:eq9} are the general form of the governing equations of plates. Its linear elasticity special case is the well-known Föppl--von Kármán (FvK) equations, named after August Föppl \cite{Foppl1899Vorlesungen} and Theodore von Kármán \cite{Karman1907Festigkeitsprobleme}. To derive them, the constitutive equations of an isotropic elastic plate should be established,

    \begin{equation}
		\label{eq:eq10}
		N_{\alpha\beta} = C\left[(1-\nu)\varepsilon_{\alpha\beta}^0 + \nu\varepsilon_{\gamma\gamma}^0\delta_{\alpha\beta}\right],
	\end{equation}
	
    \begin{equation}
		\label{eq:eq10-2}
		M_{\alpha\beta} = D\left[(1-\nu)\kappa_{\alpha\beta} + \nu\kappa_{\gamma\gamma}\delta_{\alpha\beta}\right],
	\end{equation}

    \noindent where \(\delta_{\alpha\beta}\) is the Kronecker delta (\(\delta_{\alpha\beta}=1, \text{if\ } \alpha=\beta \text{\ and\  } \delta_{\alpha\beta}=0,\text{if\ } \alpha \ne \beta\)), \(C\) is the stretching stiffness, or axial rigidity, and \(D\) is the bending stiffness, or flexural rigidity,

    \begin{equation}
		\label{eq:eq11}
		C = \frac{Eh}{1 - \nu^2}, D = \frac{{Eh}^3}{12(1 - \nu^{2})}.
	\end{equation}

    Here \(E\) is the Young's modulus and \(\nu\) is the Poisson's ratio. Substituting Eq. \eqref{eq:eq2}, Eq. \eqref{eq:eq3}, Eq. \eqref{eq:eq10} and Eq. \eqref{eq:eq10-2} into Eq. \eqref{eq:eq9}, we can get the FvK equations for the isotropic elastic plate in terms of displacement.

    \subsection{Boundary conditions}\label{boundary-conditions}

    In Eq. \eqref{eq:eq9}, the two in-plane equations have second-order spatial derivatives and the out-of-plane equation has the fourth-order spatial derivatives. Therefore, a total of eight boundary conditions are required. As in the derivation using adjacent equilibrium, it is possible to identify all these BCs on each edge of the plate. But the advantage of using the energy method is that it can give not only the governing PDEs but also the complete description of the boundary conditions that leads to a unique solution. By setting the term of the integration along the stress boundary (the second integration along \(\Gamma_{\sigma}\)) in Eq. \eqref{eq:eq8} to zero, we can obtain the boundary conditions. The quantities with a variation are referred to as the primary variables that constitute the geometric boundary conditions and the coefficients of the variations are referred to as the secondary variables that constitute the natural boundary conditions. We can see there are five primary variables \(u^0_x,\ u^0_y,\ w,\ w_{,x},\) and \(w_{,y}\) for a plate with the edges aligned with the \(x\) and \(y\) axis, which indicates a total of ten boundary conditions (five geometric and five natural boundary conditions). This is seemingly inconsistent with the eight boundary conditions from the order analysis of the PDEs. The reason is that there are only four independent primary variables among the above five variables. Only the rotation about the normal axis is considered in the plate theory. Interested readers are referred to \ref{Appendix A.3} for the detailed derivation. After a transformation from the global Cartesian coordinate to the local coordinate of \((n,\ s,\ z)\), the second integration term becomes

    \begin{equation}
		\label{eq:eq12}
		0 = \int_{\Gamma_{\sigma}}^{}\left\lbrack \left( N_{nn} - {\widehat{N}}_{nn} \right)\updelta u_{0n} + \left( N_{ns} - {\widehat{N}}_{ns} \right)\updelta u_{0s} + \left( V_{n} - {\widehat{V}}_{n} \right)\updelta w + \left( M_{nn} - {\widehat{M}}_{nn} \right) \updelta w_{,n} \right\rbrack \text{d}s,
	\end{equation}

    \noindent where

    \begin{equation}
		\label{eq:eq13}
		\begin{aligned}
		V_{n} &= \left( N_{xx}w_{,x} + N_{xy}w_{,y} \right)n_{x} + \left( N_{yy}w_{,y} + N_{xy}w_{,x} \right)n_{y} \\
		 &+ N_{xx,x}n_{x} + M_{yy,y}n_{y} + M_{xy,x}n_{y} + M_{xy,y}n_{x} + M_{ns,s}.
		 \end{aligned}
	\end{equation}

    It is then clear that the four primary variables are \(u_{0n}\), \(u_{0s}\), \(w\), and \(w_{,n}\), respectively corresponding to the in-plane displacement in the normal direction, the in-plane displacement in the tangential direction, the out-of-plane deflection, and the rotation along the normal axis, and the four secondary variables are \(N_{nn}\), \(N_{ns}\), \(V_{n}\), and \(M_{nn}\), respectively corresponding to the in-plane normal force, the in-plane tangential force, the shear force, and the bending moment.

\section{Computational framework based on ANN}\label{physics-guided-machine-learning-framework}

\begin{figure}
    \centering
    \includegraphics[width=\columnwidth]{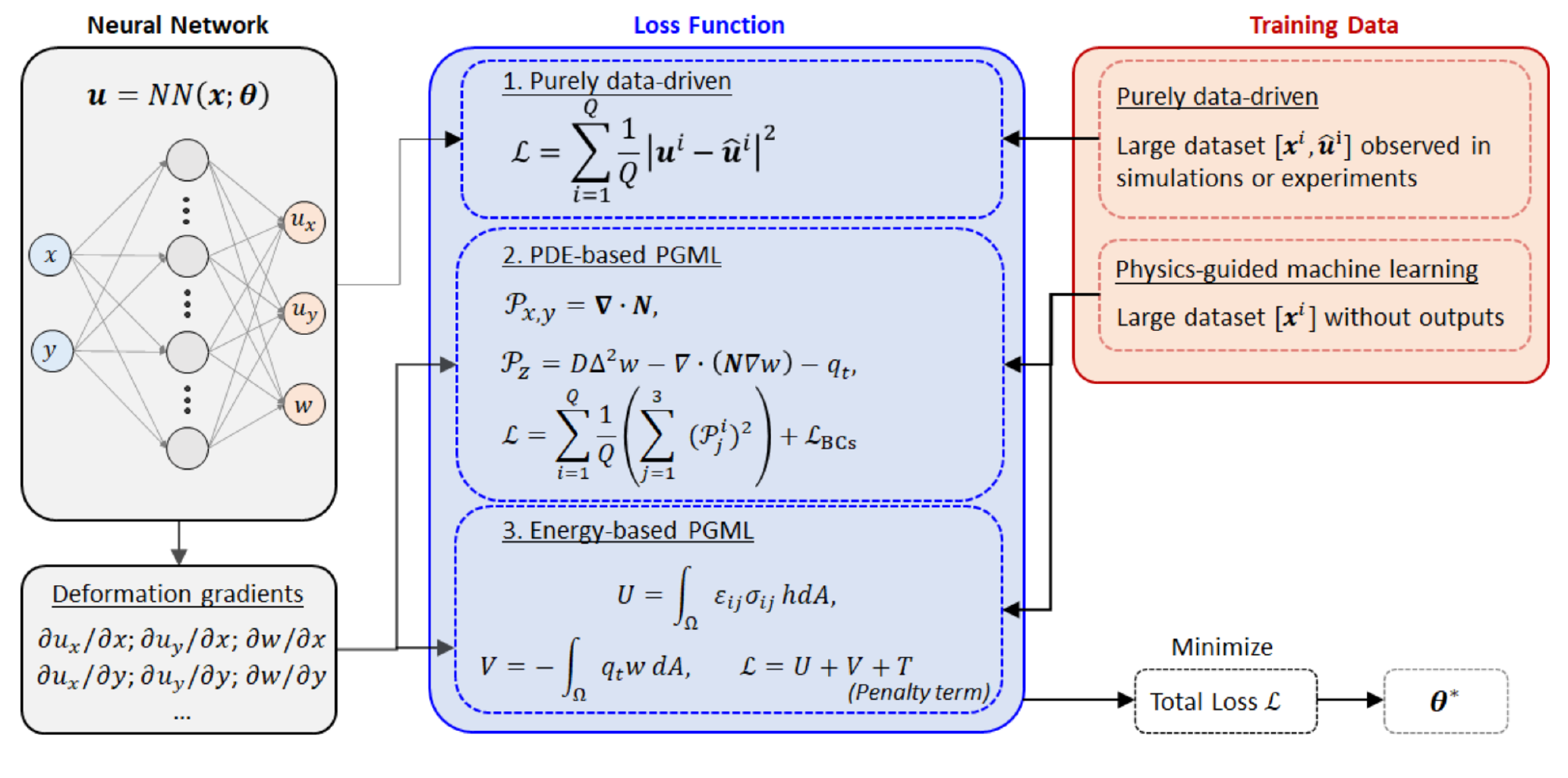}
    \caption{Flow chart of the physic-guided machine learning framework compared with the purely data-driven approach}
    \label{fig:fig2}
\end{figure}

The FvK equations are notoriously difficult to solve. Existing successful methods include semi-analytical solutions \cite{Zhu2018Stretch-induced,Cerda2002Thin,Puntel2011Wrinkling} and numerical simulations \cite{Cerda2002Thin,Nayyar2011Stretch-induced,Sipos2016Disappearance}. The former relies on making reasonable simplifications to make the equations solvable, and the latter commonly makes use of finite element (FE) methods. It should be noted that both methods can only approximate the exact solution instead of directly solving the FvK equations. In this section, we explore an alternative possible method by developing a learning framework. The overall flowchart of the algorithm is illustrated in Figure~\ref{fig:fig2}. In general, the algorithm consists of three parts, an artificial neural network, a loss function, and a training dataset, which will respectively be elaborated on in the following three sub-sections.

    \subsection{Artificial neural network}\label{artificial-neural-network}

\begin{figure}
        \centering
        \includegraphics[width=0.8\columnwidth]{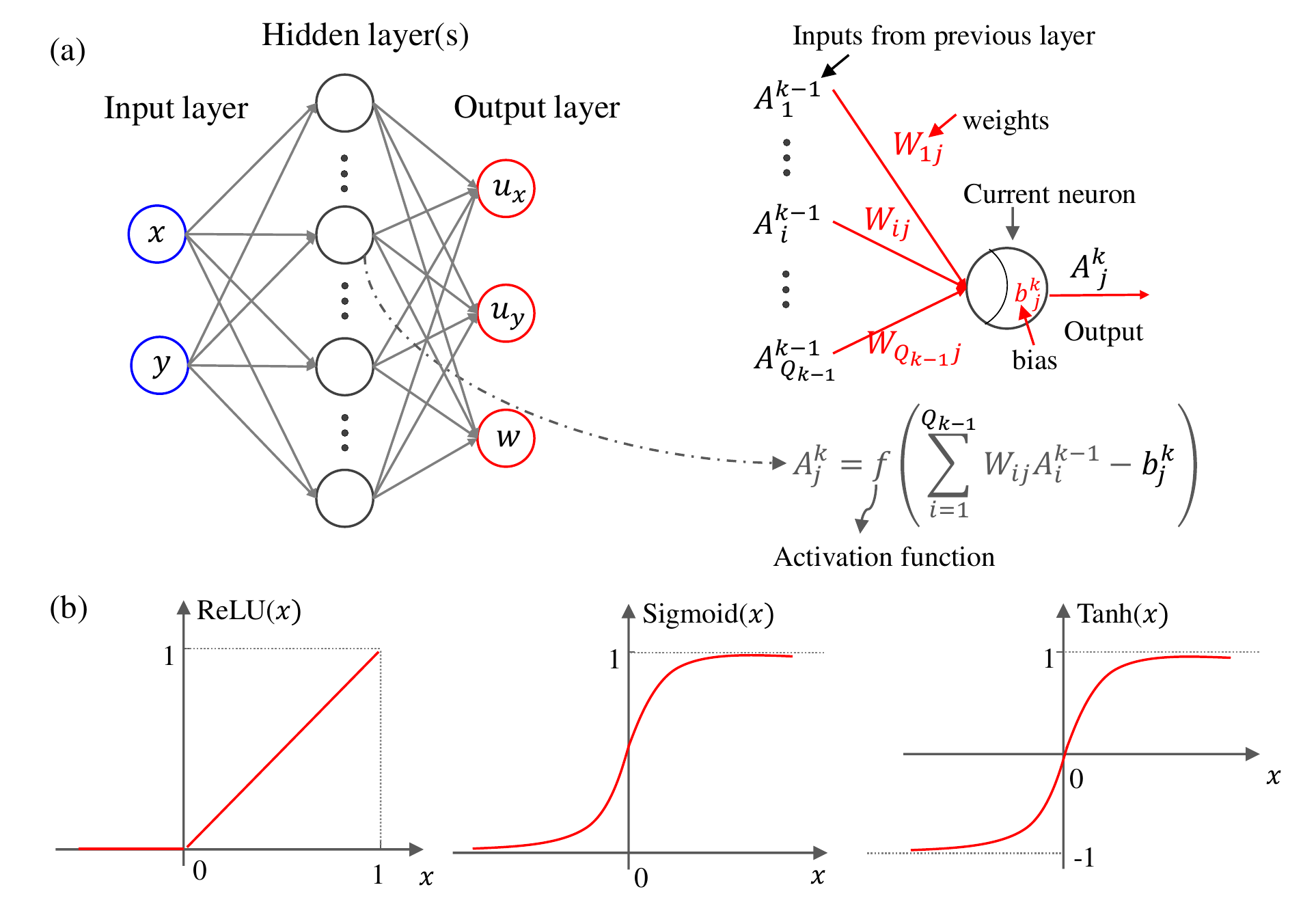}
        \caption{ Construction of the artificial neural network, (a) fully-connected multi-layer network, (b) activation functions}
        \label{fig:fig3}
    \end{figure}

    A fully connected neural network is constructed to approximate the exact solution of the displacement field. It consists of the input layer, output layer, and hidden layers in between. Here, we take the spatial coordinates \((x,y)\) as the inputs and the displacement fields \((u_x,u_y,w)\) to be predicted as the outputs (see Figure~\ref{fig:fig3}a).

    Considering an \(L\)-layer neural network, or a \((L - 1)\)-hidden layer neural network, with \(P_k\) neurons in the \(k\)-th layer (\(P_{0} = 2\) is the dimension of inputs and \(P_{L} = 3\) is the dimension of outputs), for the \(j\)-th neuron in the \(k\)-th layer, the output \(A_{j}^{k}\) is obtained by taking the weighted average of the outputs of the previous layer and then applying an activation function,

    \begin{equation}
        \label{eq:2-1}
        A_{j}^{k} = f\left( \sum_{i = 1}^{P_{k - 1}}{W_{ij}^{k}A_{i}^{k - 1}} + b_{j}^{k} \right),
    \end{equation}

    \noindent where \(W_{ij}^k\) is the weights and \(b_{j}^{k}\) is the bias. \(f(\cdot)\) represent the nonlinear activation function. Some common choices (see Fig. 3b) are the rectified linear unit (ReLU, \(f\left( x \right) = max\{x,0\}\)), the logistic sigmoid function (\(f(x) = 1/(1 + e^{- x})\)), and the hyperbolic tangent (Tanh, \(f(x) = (e^{x} - e^{- x})/(e^{x} + e^{- x})\)). The above equation can be written in vector and matrix form,

    \begin{equation}
        \label{eq:2-2}
        \mathbf{A}^{k} = f\left( \mathbf{W^k}^{T}\mathbf{A}^{k - 1} + \mathbf{b}^{k} \right),
    \end{equation}

    \noindent where \( \mathbf{W}^k=[W_{ij}^k] \in \mathbb{R}^{M_{l - 1} \times M_{l}}\) and \(\mathbf{b}^{k}=\left\lbrack b_j^{k} \right\rbrack \in \mathbb{R}^{M_{l}}\) are the weight matrix and bias vector, respectively. \(\mathbf{A}^{k}\) represents the output vector of \(k\)-th layer. The neural network can then be defined recursively as follows,

    \begin{equation}
        \label{eq:2-3}
        \begin{aligned}
            \text{input\ layer}&:\ \mathbf{A}^{0} = \left\lbrack x,y \right\rbrack \in \mathbb{R}^{2}, \\
            \text{hidden\ layers}&:\ \mathbf{A}^{k} = f\left( \mathbf{W^k}^{T}\mathbf{A}^{k - 1} + \mathbf{b}^{k} \right) \in \mathbb{R}^{M_{l}},\ \text{for}\ 1 \leq k \leq L - 1, \\
            \text{output\ layer}&:\ \mathbf{A}^{L} = \left\lbrack u,u_y,w \right\rbrack = \mathbf{W^k}^{T}\mathbf{A}^{k - 1} + \mathbf{b}^{k} \in \mathbb{R}^{3}.
        \end{aligned}
    \end{equation}

    Note that the activation function is not applied for the output layer. It should also be pointed out that the lowercase \(w\) without subscripts denotes the out-of-plane displacement whereas the capital \(\mathbf{W}\) and \(W_{ij}\) represent the weights of the neural network. The weights and biases are the parameters to be trained of the neural network and there is a total of \(\sum_{i = 1}^{L}{P_{i - 1}P_{i}}\) weights and \(\sum_{i = 1}^{L}P_{i}\) biases (\(P_{0} = 2,\ P_{L} = 3\)). It has been proven that a neural network can approximate any targeted functions with an arbitrary width or depth\cite{cybenko1989approximation}. This approximation ability of the neural network makes it possible to represent the full filed solution of the PDEs.

    \subsection{Loss function }\label{loss-function}

    The loss function is perhaps the most essential part of a neural network algorithm because the already-known physical laws will be implemented here. As illustrated in Figure~\ref{fig:fig2}, we consider three ways of formulating the loss function. The first is purely data-driven and compares the prediction of the displacement field only with the data observed from experiments or simulations. Therefore, it is not a physics-guided algorithm. We include it here for the comparison with the other two loss functions, which reflect the physical laws. The second one is defined on the PDEs and BCs, and the third one is defined on the total potential energy of the plate. It is worth noting that the PDE-based is close to the concept of PINN proposed by Raissi et al. \cite{Raissi2019Physics-informed} and the energy-based can be viewed as an implementation of the Ritz method \cite{Weinan2018Deep}. 

    \emph{\textbf{Purely data-driven}}

    Like most conventional data-driven methods, the loss function of a regression problem can be constructed by the difference between the predictions and the true experimental or numerical observations, such as the mean square error,

    \begin{equation}
        \label{eq:2-4}
        \mathcal{L}_{\text{Data-driven}} = \sum_{i = 1}^{Q}{\frac{1}{Q}\left\lbrack \left( u_x^{i} - \hat{u}_x^i \right)^{2} + \left( u_y^{i} - \hat{u}_y^{i} \right)^{2} + \left( w^{i} - {\hat{w}}^{i} \right)^{2} \right\rbrack},
    \end{equation}

    \noindent where the superscript \(i\) indicates the \(i\)-th training sample, \(\hat{u}_x^i\), \(\ \hat{u}_y^i\), and \({\hat{w}}^{i}\) are the observed displacements in the training dataset, and \(u_x^{i}\), \(\ u_y^{i}\), and \(w^{i}\) are the predicted displacements, \(Q\) is the total number of training samples, which should be sufficient large to ensure an acceptable accuracy.
    
    It is worth noting that the purely data-driven loss function can also be defined on the strain field or the stress field. To realize these formulations, one can replace the {\(u, w\)} and {\(\hat{u},\hat{w}\)} components in Eq.{\eqref{eq:2-4}}. It is also common to use a combination of the displacement, strain, and stress fields. This point will be investigated in the following sections of the paper.

    \emph{\textbf{PDE-based}}

    For the studied plate theory, the outputs of the neural network should satisfy the governing PDEs (Eq. \eqref{eq:eq9}) and the BCs (Eq. \eqref{eq:eq12}). The first way to implement the plate theory into the algorithm is, therefore, to construct the loss with the residuals of the PDEs and BCs,

    \begin{equation}
        \label{eq:2-5}
        \mathcal{L}_{\text{PDE-based}} = \mathcal{L}_{\text{PDEs}} + \lambda_{s}\mathcal{L}_{\text{BCs}} + \lambda_{d}\mathcal{L}_{\text{BCd}},
    \end{equation}

    \noindent where \(\mathcal{L}_{\text{PDEs}}\) is the residual of the PDEs,

    \begin{equation}
        \label{eq:2-6}
        \mathcal{L}_{\text{PDEs}} = \sum_{i = 1}^{Q_{\text{P}}}{\frac{1}{Q_{\text{P}}}\left\lbrack \left( \mathcal{P}_{x}^{i} \right)^{2} + \left( \mathcal{P}_{y}^{i} \right)^{2} + \left( \mathcal{P}_{z}^{i} \right)^{2} \right\rbrack},
    \end{equation}

    \(\mathcal{P}_{x,y,z}\) are the residual values of the three governing PDEs defined in Eq. \eqref{eq:eq9}, \(Q_{\text{P}}\) is the number of training samples within the solution domain.

    \(\mathcal{L}_{\text{BCs}}\) is the residual of the static BCs,

    \begin{equation}
        \label{eq:2-7}
        \mathcal{L}_{\text{BCs}} = \sum_{i = 1}^{Q_{\text{bs}}}{\frac{1}{Q_{\text{bs}}}\left\lbrack \left( N_{nn}^{i} - {\widehat{N}}_{nn}^{i} \right)^{2} + \left( N_{ns}^{i} - {\widehat{N}}_{ns}^{i} \right)^{2} + \left( V_{n}^{i} - {\hat{V}}_{n}^{i} \right)^{2} + \left( M_{nn}^{i} - {\widehat{M}}_{nn}^{i} \right)^{2} \right\rbrack},
    \end{equation}

    \(\mathcal{L}_{\text{BCd}}\) is the residual of the kinematic BCs,

    \begin{equation}
        \label{eq:2-8}
        \mathcal{L}_{\text{BCd}} = \sum_{i = 1}^{Q_{\text{bd}}}{\frac{1}{Q_{\text{bd}}}\left\lbrack \left( u_{0n}^{i} - {\hat{u}}_{0n}^{i} \right)^{2} + \left( u_{0s}^{i} - {\hat{u}}_{0s}^{i} \right)^{2} + \left( w^{i} - \hat{w}^{i} \right)^{2} + \left( w^i_{,n} - {{\hat{w}}_{,n}}^{i} \right)^{2} \right\rbrack}.
    \end{equation}

    \(\lambda_{s}\) and \(\lambda_{d}\) are the weights of loss on stress boundary and displacement boundary, the hat notation indicates the prescribed secondary and primary variables defined in Eq. \eqref{eq:eq12} at the boundaries, and the corresponding variables without a hat are the predicted outputs of the neural network, \(Q_{\text{bs}}\) and \(Q_{\text{bd}}\) are the number of samples at the static boundary and kinematic boundary, respectively.

    By minimizing the total loss, the PDEs and boundary conditions can be satisfied. Mathematically, the neural network is able to approximate the exact solution.

    \emph{\textbf{Energy-based}}

    The second way to implement the plate theory into the algorithm is to directly take the total potential energy as the loss and minimize the loss according to the principle of minimum potential energy. However, as E et al. \cite{Weinan2018Deep} pointed out in their study on the deep Ritz method, the challenging issue is how to incorporate the kinematic boundaries into the total potential energy because it is not automatically included. Here, we construct the loss function based on the total potential energy with a penalty energy term \cite{Weinan2018Deep,Zhu2018Stretch-induced,Cerda2002Thin},

    \begin{equation}
        \label{eq:2-9}
        \mathcal{L}_{\text{Energy-based}}=\Pi = U+V+T,
    \end{equation}

    \noindent where \(U\) is the internal energy,

    \begin{equation}
        \label{eq:2-10}
        U = \int_{\Omega}^{}{\frac{1}{2}\left( N_{\alpha\beta}\varepsilon_{\alpha\beta}^{0} +  M_{\alpha\beta}\kappa_{\alpha\beta} \right) \text{d}x\text{d}y},
    \end{equation}

    \noindent \(V\) is the virtual work done by external forces on the boundary (defined as negative to be consistent with the principle virtual displacement)

    \begin{equation}
        \label{eq:2-11}
        V = \int_{\Gamma_{\sigma}}^{}{\left\lbrack - {\widehat{N}}_{nn}u_{0n} - {\widehat{N}}_{ns}u_{0s} - {\widehat{N}}_{nz}w + {\widehat{M}}_{ns}w_{,s} + {\widehat{M}}_{nn} w_{,n} \right\rbrack \text{d}s},
    \end{equation}

    \noindent and \(T\) is the penalty term that enforces the kinematic boundary conditions,

    \begin{equation}
        \label{eq:2-12}
        T = \lambda_{n}^{*}\int_{\Gamma_{d}}^{}{\left| u_{0n} - {\hat{u}}_{0n} \right|\text{d}s} + \lambda_{s}^{*}\int_{\Gamma_{d}}^{}{\left| u_{0s} - {\hat{u}}_{0s} \right|\text{d}s} + \lambda_{w}^{*}\int_{\Gamma_{d}}^{}{\left| w - \hat{w} \right|\text{d}s} + \lambda_{w,n}^{*}\int_{\Gamma_{d}}^{}{\left| w_{,n} - {\hat{w}}_{,n} \right|\text{d}s}.
    \end{equation}

    Here, $\lambda_{n}^{*}$, $\lambda_{s}^{*}$, $\lambda_{w}^{*}$, and $\lambda_{w,n}^{*}$ are four Lagrangian multipliers that are in the dimension of force ($\lambda_{n}^{*}$, $\lambda_{s}^{*}$, $\lambda_{w}^{*}$) or moment ($\lambda_{w,n}^{*}$), representing the applied forces and moments on the kinematic boundaries. Note that if the kinematic BCs are satisfied, this penalty term vanishes. In addition, the first variation of this penalty term is zero (\(\updelta T = 0\)), and therefore, taking the variation of the total potential energy functional \(\Pi\) still returns \(\updelta U + \updelta V\). This is consistent with Eq. \eqref{eq:eq1}.

    The energy integrations can be evaluated numerically. For integration in the domain, we first uniformly sample \(Q_{\text{P}}\) points, the integration can then be approximated by,

    \begin{equation}
        \label{eq:2-13}
        U_{\text{num}} = \sum_{i = 1}^{Q_{\text{P}}}{\overline{U}^{i}\updelta A_{i}} \approx \sum_{i = 1}^{Q_{\text{P}}}{\frac{1}{Q_{\text{P}}}\overline{U}^{i}A_{t}},
    \end{equation}

    \noindent where \(\overline{U}\) is internal energy density

    \begin{equation}
        \label{eq:2-13-2}
        \overline{U}=\frac{1}{2}\left(N_{\alpha\beta}\varepsilon_{\alpha\beta}^{0} + M_{\alpha\beta}\kappa_{\alpha\beta} \right).
    \end{equation}

    \noindent \(\overline{U}^{i}\) is, therefore, the internal energy density of the \(i\)-th sample point, \(\updelta A_{i}\) is the discrete area around the sampled point, which can be estimated by \(A_{t}/Q_{\text{P}}\) is the total area when \(Q_{\text{P}}\) is sufficiently large.

    Similarly,

    \begin{equation}
        \label{eq:2-14}
        V_{\text{num}} = \sum_{i=1}^{Q_{\text{bs}}}{\overline{V}^{i}\updelta l_{i}} \approx \sum_{i = 1}^{Q_{\text{bs}}}{\frac{1}{Q_{\text{bs}}}\overline{V}^{i}l_{t}},
    \end{equation}

    \noindent where \(\overline{V}\) is the external work per unit width,

    \begin{equation}
        \label{eq:2-15}
        \overline{V} = -{\widehat{N}}_{\text{nn}}u_{0n} - {\widehat{N}}_{\text{nz}}w - {\widehat{N}}_{\text{ns}}u_{0s} + {\widehat{M}}_{\text{ns}} w_s + {\widehat{M}}_{\text{nn}} w_n.
    \end{equation}

    \noindent \(\updelta l_{i}\) is the discrete length around the sampled point, approximately \(l_{t}/Q_{\text{BCs}}\), where \(l_{t}\) is the total length of the static boundaries.

    \(T\) can be numerically interpreted in the same way as \(\mathcal{L}_{\text{BCd}}\) in Eq. \eqref{eq:2-13}. For simplicity, we assume all the four Lagrangian multipliers are of the same value $\lambda_{d}$,

    \begin{equation}
        \label{eq:2-16}
        T_\text{num}=\lambda_{d}\sum_{i = 1}^{Q_{\text{bd}}}{\frac{1}{Q_{\text{bd}}}\left\lbrack \left( u_{0n}^{i} - {\hat{u}}_{0n}^{i} \right)^{2} + \left( u_{0s}^{i} - {\hat{u}}_{0s}^{i} \right)^{2} + \left( w^{i} - {\hat{w}}^{i} \right)^{2} + \left( w_{,n}^{i} - {\hat{w}}_{,n}^{i} \right)^{2} \right\rbrack}^{1/2}.
    \end{equation}

    It should be noted that both PDE-based and energy-based loss functions require the calculation of the partial derivatives of the outputs with respect to the inputs. Most existing machine learning platforms (e.g. Tensorflow \cite{abadi2016tensorflow} and Pytorch \cite{paszke2017automatic}) are already equipped with default gradient algorithms, and users can obtain the numerical gradient efficiently. Higher-order partial derivatives can also be calculated by the algorithms but take more computation time. An alternative strategy is to introduce the derivatives into the outputs of the neural network. For example, if we set the outputs to be \(( u_x,\ u_y,\ u_{x,x},\ u_{x,y},\ u_{y,x},\ u_{y,y})\), the second-order derivatives of \(u_x\) and \(u_y\) can be obtained by performing only the first derivative of the outputs. In this way, we can avoid calculating high-order derivatives. The disadvantage is that it will increase the size of the neural network and extra constraints are needed to enforce the mathematical relations among the outputs. For instance, in the above example, the derivative of the first output \(u_x\) to the first input \(x\) should be equal to the third output \(u_{x,x}\). For simplicity, we adopted the default gradient algorithm in Pytorch to ensure higher accuracy of the calculation of the derivatives at the sacrifice of computational efficiency.

    \subsection{Training dataset}\label{training-dataset}

    As indicated by Eq. \eqref{eq:2-4}, the purely data-driven model loss function can be minimized only when the displacement field \(\left( \hat{u}_x,\hat{u}_y,\hat{w} \right)\) of a sufficiently large number of sample points can be observed. It means a large experimental or numerical database. The exact solution of \(\left( \hat{u}_x,\hat{u}_y,\hat{w} \right)\) is always preferable rather than numerical simulation results for a reliable training process. Experimental data is good but usually comes with measurement uncertainties. In addition, as we already pointed out, high-order equations such as the FvK equations are difficult to be solved analytically. Therefore, how to obtain a satisfactory dataset to train the data-driven algorithm is a big challenge even for a simple structure like an elastic plate.

    The training data of the physics-guided algorithms has two important features: 1) exact solutions are not required (the loss function only contains the predicted outputs). It means that the training data are simply the input spatial coordinates. 2) it should be sampled both within the solution domain and at the boundaries. Theoretically, the training data can be sampled in any size and strategy. Two possible sampling strategies are: 1) the data points are sampled in the beginning and remain unchanged during training process. The samples can be uniform grid points or randomly distributed points. 2) the data points are sampled randomly at each training epoch, which means the training dataset changes during training. For energy-based loss function, due to the requirement of numerical integration (Eq. \eqref{eq:2-13} and Eq. {\eqref{eq:2-14}}), we choose the second strategy where we first randomly sample the data points in a uniform distribution and then resample the data during each training epoch to have more accurate and consistent results. For the PDE-based loss function, the same strategy is used. It should also be noted that the data points can be non-uniformly distributed with local refinement, which can be used as a means to improve the results. This point will be further discussed in the next section.
    
    The training process is essentially optimizing the weights and bias of a neural network to minimize the loss function. It is common to train the network with small batches from the training dataset. This is still applicable for PDE-based loss function. For the energy-based loss function, however, the network has to be trained with one batch (the whole dataset) for each iteration of optimization to reliably evaluate the numerical integration.

\section{Applications and validations}\label{applications-and-validations}

To validate the machine learning framework that we developed, four typical loading conditions will be characterized, and the prediction will be compared with the FE simulation results. Since the geometry of the specimen and boundary conditions are not complex, the FE simulations can approximately represent the exact solution.

\subsection{In-plane tension with non-uniformly distributed stretching force}\label{in-plane-tension-with-non-uniformly-distributed-stretching-force}

\begin{figure}
        \centering
        \includegraphics[width=0.8\columnwidth]{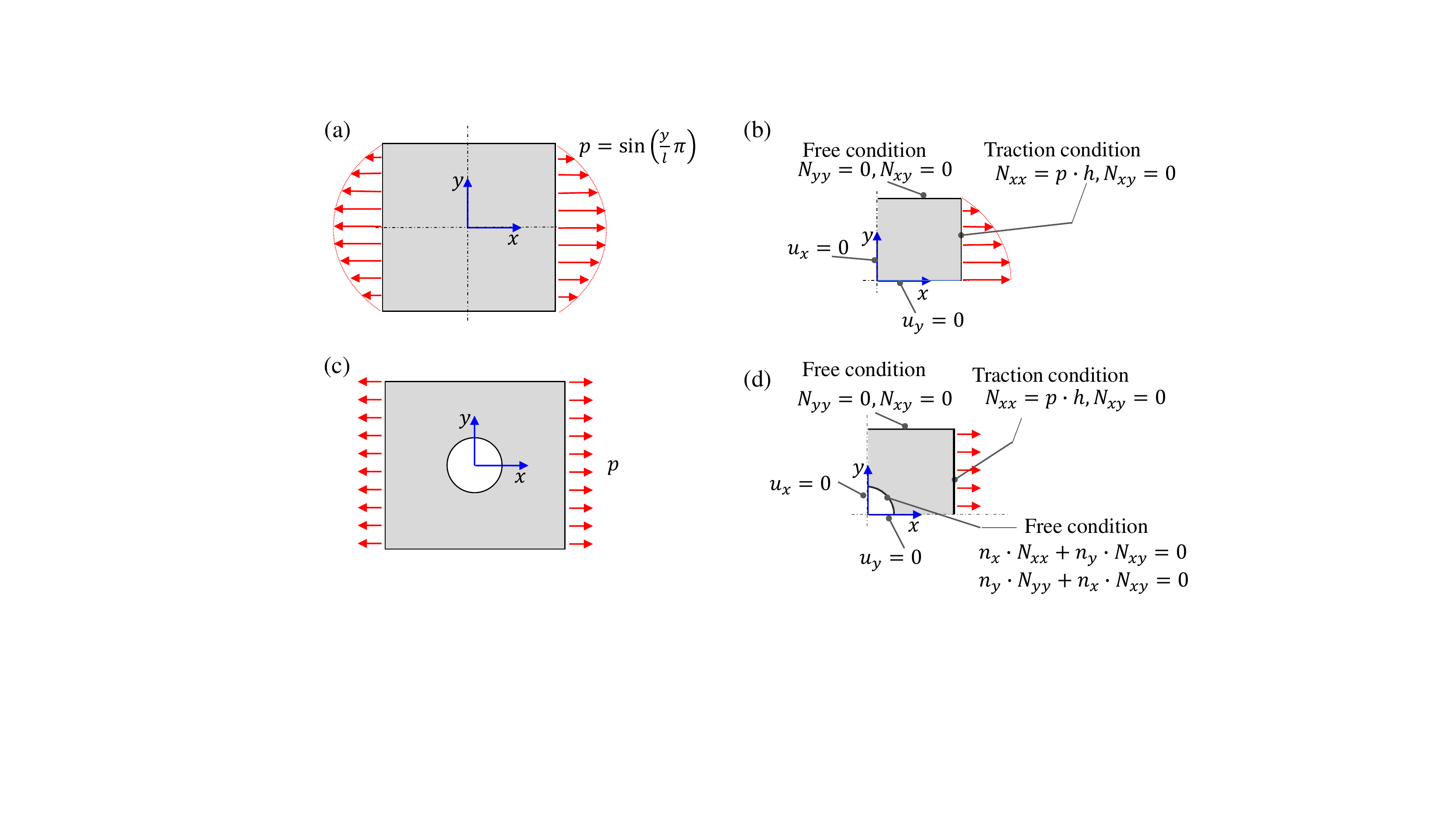}
        \caption{(a) Loading condition of non-uniform tension of plate and (b) one-quarter equivalent model;(c) loading condition of uniaxial central-hole tension and (d) one-quarter equivalent model.}
        \label{fig:fig4}
    \end{figure}

We start with a two-dimensional case that only involves in-plane deformation under plane-stress condition, as illustrated in Figure~\ref{fig:fig4}a. The two lateral edges of a 20 mm \(\times\) 20 mm (\(l \times l\)) square elastic plate are subjected a non-uniform stretching force that follows a sinuous distribution \(p = \sin\left( y\pi/l \right)\) MPa. The other two edges (upper and lower) are traction-free. The origin of the Cartesian coordinate is placed at the center of the square with the \(x\) axis pointing to the right. The Young's modulus of the elastic plate is 70 MPa and the Poisson's ratio is 0.3. Due to the symmetry of the geometry and loads, only one quarter of the plate is modeled (see Figure~\ref{fig:fig4}b). The boundary conditions at the four edges of the quarter model are listed as following,

\begin{equation}
    \label{eq:3-1}
    \begin{aligned}
        u_x|_{x = 0} &= 0, \\
        u_y|_{y = 0} &= 0, \\
        N_{xx}|_{x = \frac{l}{2}} &=p \cdot h = \sin\left( \frac{y}{l}\pi \right)h, &N_{xy}|_{x = \frac{l}{2}} = 0, \\
        N_{yy}|_{y = \frac{l}{2}} &= 0,&N_{xy}|_{y = \frac{l}{2}} = 0.
    \end{aligned}
\end{equation}

    The two governing PDEs for this case can be written in terms of the displacement field,

    \begin{equation}
        \label{eq:3-2}
        \begin{aligned}
            \mathcal{P}_{x} &\equiv \frac{E}{1 - \nu^{2}}\left( \frac{\partial^{2}u_x}{\partial x^{2}} + \frac{1 - \nu}{2}\frac{\partial^{2}u_x}{\partial y^{2}} + \frac{1 + \nu}{2}\frac{\partial^{2}u_y}{\partial x\partial y} \right) = 0,\\
            \mathcal{P}_{y} &\equiv \frac{E}{1 - \nu^{2}}\left( \frac{\partial^{2}u_y}{\partial y^{2}} + \frac{1 - \nu}{2}\frac{\partial^{2}u_y}{\partial x^{2}} + \frac{1 + \nu}{2}\frac{\partial^{2}u_x}{\partial x\partial y} \right) = 0.
        \end{aligned}
    \end{equation}

    Two different loss functions are defined for comparison according to Eq. \eqref{eq:2-5} and Eq. \eqref{eq:2-9}. The PDEs-based loss function is,

    \begin{equation}
        \label{eq:3-3}
        \mathcal{L}_{\text{PDE-based}} = \mathcal{L}_{\text{PDEs}} + \lambda_s ( \mathcal{L}_{\text{BCsx}} + \mathcal{L}_{\text{BCsx}} ) + \lambda_d \mathcal{L}_{\text{BCd}},
    \end{equation}

    and

    \begin{equation}
        \label{eq:3-4}
        \begin{aligned}
            \mathcal{L}_{\text{PDEs}} &= \frac{1}{Q_{\text{P}}}\sum_{i = 1}^{Q_{\text{P}}}\left\lbrack \left( \mathcal{P}_{x}^{i} \right)^{2} + \left( \mathcal{P}_{y}^{i} \right)^{2} \right\rbrack, \\
            \mathcal{L}_{\text{BCsx}} &= \frac{1}{Q_{\text{sx}}}\sum_{i = 1}^{Q_{\text{sx}}}\left\{ \left\lbrack \left( N_{xx}^{i} \right)^{2} - p\cdot h \right\rbrack^{2} + \left( N_{xy}^{i} \right)^{2} \right\}, \\
            \mathcal{L}_{\text{BCsy}} &= \frac{1}{Q_{\text{sy}}}\sum_{i = 1}^{Q_{\text{sy}}}\left\lbrack \left( N_{xx}^{i} \right)^{2} + \left( N_{xy}^{i} \right)^{2} \right\rbrack, \\
            \mathcal{L}_{\text{BCd}} &= \frac{1}{Q_{\text{dx}}}\sum_{i = 1}^{Q_{\text{dx}}}\left( u_x^{i} \right)^{2} + \frac{1}{Q_{\text{dy}}}\sum_{i = 1}^{Q_{\text{dy}}}\left( u_y^{i} \right)^{2},
        \end{aligned}
    \end{equation}

    \noindent where variables with superscript \(i\) indicate value evaluated for the \(i\)-th training sample, \(Q_{\text{P}}\), \(Q_{\text{sx}}\), \(Q_{\text{sy}}\ Q_{\text{dx}}\), and \(Q_{\text{dy}}\) are the total number of training samples within the domain, on the right and upper edges (static boundary), and on the left and lower edges (kinematic boundary), respectively. The membrane forces are computed with the displacement gradients according to Eq.~\ref{eq:eq10}.

    The energy-based loss function is,

    \begin{equation}
        \label{eq:3-6}
        \begin{aligned}
  			\mathcal{L}_{\text{Energy-based}} &= U +V + T \\
            & =\int_{0}^{l/2}\int_{0}^{l/2}\frac{Eh}{2\left( 1 -  u_y^{2} \right)}\bigg[ \left( \frac{\partial u_x}{\partial x} \right)^{2} + \left( \frac{\partial u_y}{\partial y} \right)^{2} + 2\nu\frac{\partial u_x}{\partial x}\frac{\partial u_y}{\partial y} \\
            & + \frac{1 - \nu}{2}\left( \frac{\partial u_x}{\partial y} + \frac{\partial u_y}{\partial x} \right)^{2} \bigg]\text{d}x\text{d}y \\
            & - \int_{0}^{l}{\left\lbrack N_{xx}(y) \cdot u_x(y) \right\rbrack_{x = l/2}\text{d}y} + \int_{0}^{l/2}{\left\lbrack u_x^{2}\left( y \right) \right\rbrack_{x = 0}h\text{d}y} 
            + \int_{0}^{l/2}{\left\lbrack u_y^{2}\left( x \right) \right\rbrack_{y = 0}h\text{d}x}.      
        \end{aligned}
    \end{equation}

    The fully connected neural network defined in Eq. \eqref{eq:2-3} is used, where the outputs are replaced by 2D displacement fields \((u_x,u_y)\). We modified the outputs in the following way,

    \begin{equation}
        \label{eq:3-7}
        \begin{aligned}
            \left\lbrack u_x',u_y' \right\rbrack &= \mathcal{N}\left( x,y \right) \\
            u_x &= u_x' \cdot x, \\
            u_y &= u_y' \cdot y,
        \end{aligned}
    \end{equation}

    \noindent where \(u_x'\) and \(u_y'\) are the outputs of neural network \(\mathcal{N}\left( x,y \right)\), \(u_x\) and \(u_y\) are the modified outputs. The displacement boundary conditions can then be satisfied (\(u_x|_{x = 0} = 0,\ u_y|_{y = 0} = 0\)), which means that the loss term \(\mathcal{L}_{\text{BCd}}\) is always zero. This can simplify the calculation of loss function (\(T \equiv 0\) is automatically satisfied).

    At the same time, we solved the problem using finite element method with an extremely fine mesh size of 0.1 mm (10,000 elements in total) in Abaqus/standard. Since this is a simple mechanical problem, the results were regarded as the extract solution to evaluate the accuracy of the machine learning methods.

\begin{figure}
        \centering
        \includegraphics[width=\columnwidth]{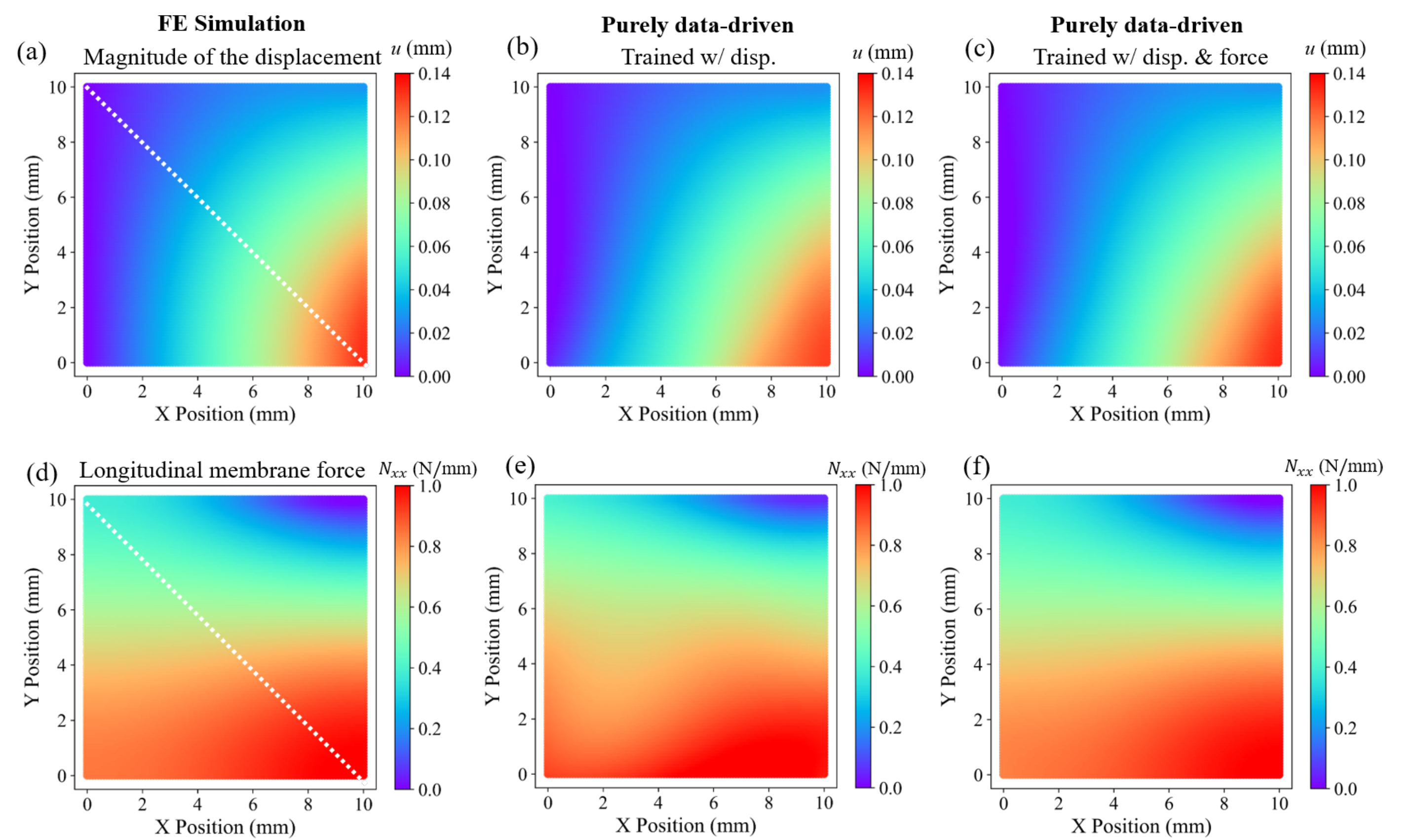}
        \caption{Predicted horizontal displacement and longitudinal membrane force of FE simulation (a,d), and purely date-driven models trained with different loss functions defined by: (b,e) mean square error of displacement field, (c,f) mean square error of displacement and membrane force fields, under non-uniform stretching load. (For interpretation of the references to color in this figure legend, the reader is referred to the web version of this article.)}
        \label{fig:fig5}
    \end{figure}

\begin{figure}
        \centering
        \includegraphics[width=\columnwidth]{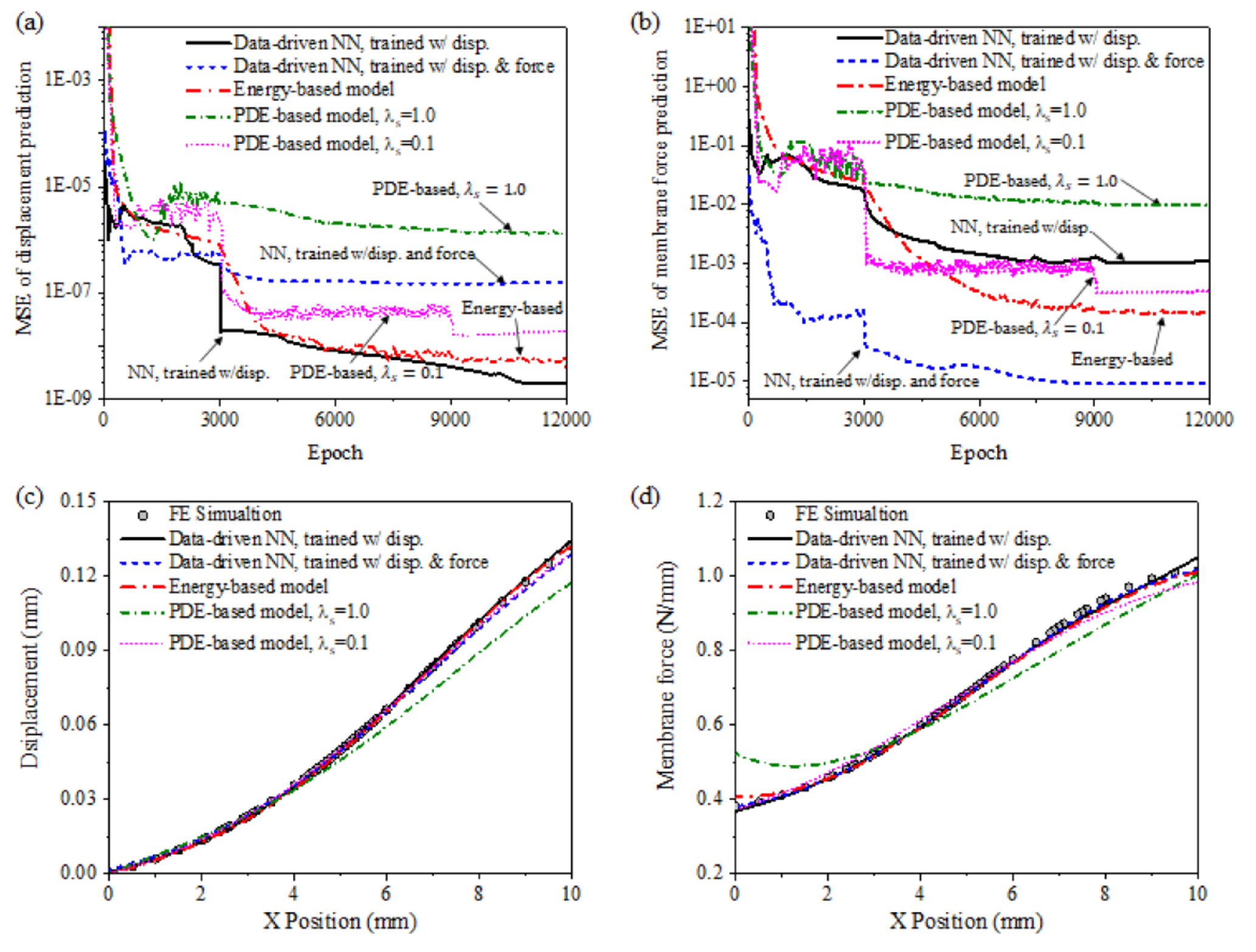}
        \caption{Mean square error of displacement (a) and membrane force (b) predictions.  Predicted horizontal displacement (c) and longitudinal membrane force (d) along the diagonal line.}
        \label{fig:fig6}
    \end{figure}

    We performed the purely data-driven machine learning method to be compared with our physics-guided methods. The training data were extracted from the FE simulation results with the spatial positions \(\left( x,y \right)\) as the inputs and the displacement fields \((\hat{u_x},\hat{u_y})\) as the outputs. 10,000 samples were used for training, with a batch size of 128 and varied training rates ($10^{-3}$ for initial 3,000 epochs, $10^{-4}$ for subsequent 6,000 epochs, and $10^{-5}$ for another 3000 epochs).  The hyperbolic (Tanh) function was used as the activation function. The loss function is constructed with the mean square error (MSE) of the displacement field. Figure~\ref{fig:fig5}a and b show a comparison between the contour plots of the magnitude of the displacement obtained by FE simulation and the predictions of a trained 5-hidden layer (5 neurons each layer) neural network. The predicted longitudinal membrane force field was then obtained according to Eq. \eqref{eq:eq10}, as shown in Figure~\ref{fig:fig5}d. The coefficients of determination (namely, $R^2$  value) of the predicted outputs were calculated to evaluate the accuracy of the global prediction of the displacement and membrane force fields (see Table 1). We can see that though the 5-hidden layer network can well predict both fields globally, the accuracy of membrane force prediction is not as high as that of displacement field (see Figure~{\ref{fig:fig5}}d and e as well as $R^2$ values in Table~{\ref{tab:tab1}}). This is mainly because the membrane force is not directly included in the loss function and the derivative operation used to calculate membrane force magnifies the error of displacement prediction. We then trained the same network with the same procedure but with a different loss function that was defined by the MSE of both displacement and membrane force fields. The results (Figure~{\ref{fig:fig5}}c and f) show that the accuracy of membrane force prediction is improved without affecting the accuracy of displacement prediction. Figure~{\ref{fig:fig6}}a and b respectively present the MSE of the displacement and membrane force versus the training epochs. The final MSE of membrane force is significantly decreased when the network is trained with both fields. To further visualize the accuracy locally, we plotted the displacement and membrane force along the diagonal line of the square plate. As shown in Figure~{\ref{fig:fig6}}c and d, the same phenomenon can be seen.

    \begin{table}
        \centering
        \caption{Coefficients of determination of the predicted displacement and membrane force fields under non-uniform stretching load}
        \label{tab:tab1}
        \begin{tabular}{l|c|c|c|c|c|}
            \toprule
            Neural network     & \multicolumn{5}{c}{Coefficient of determination $(R^2)$} \\
                               & \(u_x\) & \(u_y\) & \(N_{xx}\) & \(N_{yy}\) & \(N_{xy}\) \\ \midrule
            Data-driven (w/ disp.)  & 0.9993 &	0.9977 &	0.9959	& 0.9341	& 0.7101  \\
            Data-driven (w/ disp. and force)  & 0.9984 &	0.9904 &	0.9990	& 0.9993	& 0.9925 \\
            PDE-based \((\lambda_s=1.0)\) & 0.9602 &	0.8851 &	0.9241	& 0.8665	& 0.7688 \\
            PDE-based \((\lambda_s=0.1)\) & 0.9993 &	0.9816 &	0.9911	& 0.9645	& 0.8413 \\
            PDE-based (1,000 samples)    & 0.9988 &	0.9858 &	0.9942	& 0.9789	& 0.8415 \\
            PDE-based (200 samples)    & 0.9985 &	0.9885 &	0.9914	& 0.9719	& 0.8231 \\
            Energy-based  &0.9986 &	0.9993 &	0.9943	& 0.9900	& 0.9032\\
            Energy-based  (1,000 samples) & 0.9869 &	0.9782 &	0.9497	& 0.9469	& 0.7072
            \\
            Energy-based  (200 samples) & 0.9603 &	0.8715 &	0.7822	& 0.3245	& 0.1234\\
            \bottomrule
        \end{tabular}
    \end{table}

\begin{figure}
        \centering
        \includegraphics[width=\columnwidth]{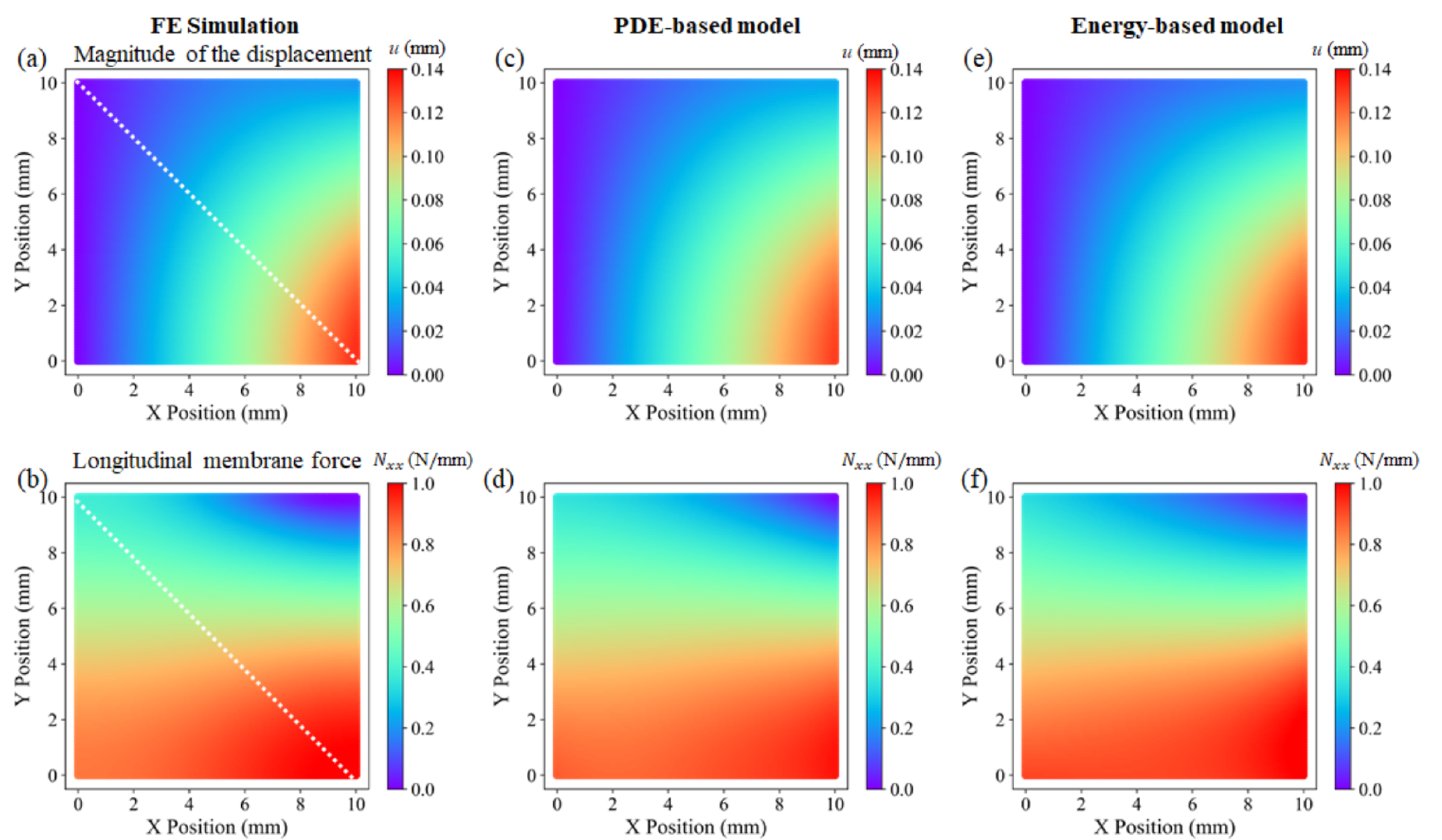}
        \caption{Predicted horizontal displacement and longitudinal membrane force of FE simulation (a,b), PDE-based model with $\lambda_s=0.1$ (c,d), and energy-based model (e,f) under non-uniform stretching load. (For interpretation of the references to color in this figure legend, the reader is referred to the web version of this article.)}
        \label{fig:fig7}
    \end{figure}

    The above analysis provides a general idea on how well the ANN can predict the displacement field. For the models with physics-guided loss functions, we use the same size of the network (5 layers and 5 neurons each layer). We uniformly sampled 10,000 data points within the solution area and 1,000 data points from each edge. The model is first trained with the PDE-based loss function defined in Eq. \eqref{eq:3-3} in the procedure described in Section~{\ref{physics-guided-machine-learning-framework}}. The hyperparameter $\lambda_s$ in Eq.{\eqref{eq:3-3}}, the weight of BC residuals, is carefully tuned. Here, we show the results of two cases with $\lambda_s=1.0$ and $\lambda_s=0.1$ as a comparison. The latter turns out to provide a better performance (see the MSE in Figure~{\ref{fig:fig6}} and $R^2$ values in Table~{\ref{tab:tab1}}). The predicted displacement and membrane force fields are shown in Figure~{\ref{fig:fig7}}c and d, where the distribution is well captured with a high accuracy. The model is then trained with the energy-based loss function, the accuracy of both displacement and membrane force field is similar to that of the PDE-based loss function. In terms of the computational efficiency, the energy-based model takes much shorter time (20 minutes for energy-based and 160 minutes for PDE-based, with 4-core Intel Core I5 CPU without GPU acceleration). This is largely because the calculation of the high-order derivatives are not required in the energy-based method.
    
\begin{figure}
        \centering
        \includegraphics[width=\columnwidth]{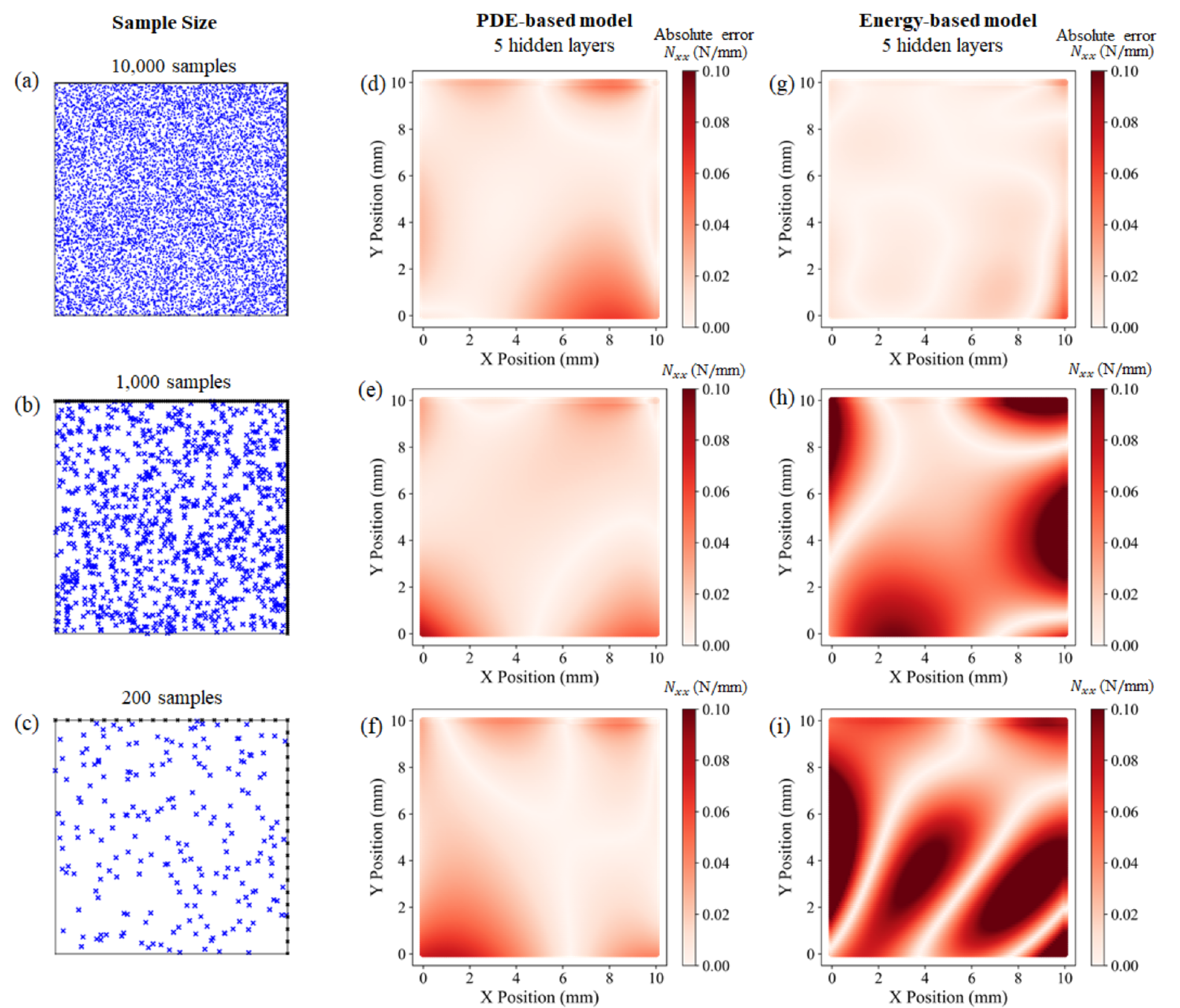}
        \caption{Absolute error of predicted longitudinal membrane force of PDE-based and energy-based models trained with different size of training samples: (a, d, g) 10,000 samples, (b, e, h) 1,000 samples, (c, f, i) 200 samples. (For interpretation of the references to color in this figure legend, the reader is referred to the web version of this article.)}
        \label{fig:fig8}
    \end{figure}

    \noindent \textit{Discussion on sampling size}. For the purely data-driven methods, the accuracy highly relies on the quality and size of the training data. To investigate the influence of the sampling size on the two physics-guided models (energy-based and PDE-based), we trained the models with two smaller sampling size (1,000 and 200 samples within the domain and 100 and 20 on the edge, respectively), as shown in Figure~{\ref{fig:fig8}}b and c. The absolute error of membrane force prediction with different sampling size is shown in Figure~{\ref{fig:fig8}}d, e, and f for the PDE-based model and Figure~{\ref{fig:fig8}}g, h, and i for the energy-based model. The accuracy of energy-based model significantly decreases with a smaller sample size, but the PDE-based model still has a relatively high accuracy. This disadvantage will be magnified in the next example where stress concentration exists in the specimen. The underlying mechanism will be further discussed in Sections~{\ref{discussion_comparison_PDE_Energy}} and {\ref{discussion_difference}}.

    \subsection{In-plane central-hole tension}\label{in-plane-central-hole-tension}

    The second application still focuses on the in-plane loading, but we will carefully investigate the effect of geometric nonlinearity by introducing a hole at the center of the specimen. Thus, the governing equations are the same as Eq. \eqref{eq:3-2}. The central hole leads to the stress concentration phenomenon at the edge of the hole, as a tough loading case for the machine learning method to solve. The problem is illustrated in Figure~\ref{fig:fig4}c, the hole has a diameter of 5 mm in the center of 20 mm \(\times\) 20 mm (\(l \times l\)) square plate. A uniformly distributed stretching load (1 MPa) is applied at both ends. The loading case is equivalent to the one-quarter model shown in Figure~\ref{fig:fig4}d and the boundary conditions for the 4 straight edges and 1 arch edge are listed as follows,

    \begin{equation}
        \label{eq:3-8}
        \begin{aligned}
            {u_x|}_{x = 0} &= 0, \\
            u_y|_{y = 0} &= 0, \\
            N_{xx}|_{x = \frac{l}{2}} &= p \cdot h,\ N_{xy}|_{x = \frac{l}{2}} = 0, \\
            N_{yy}|_{y = \frac{l}{2}} &= 0,\  N_{xy}|_{y=\frac{l}{2}} = 0,\\
            N_{xx} n_x + N_{xy} n_y &= 0,\ N_{xy} n_x + N_{xx} n_y = 0, (\text{at}x^2 + y^2={d^2}/{4}),
        \end{aligned}
    \end{equation}
    \noindent where $n_x$ and $n_y$ are the direction cosines of the boundary normal vector as defined in Eq.~\ref{eq:5-8}.
    
\begin{figure}
        \centering
        \includegraphics[width=\columnwidth]{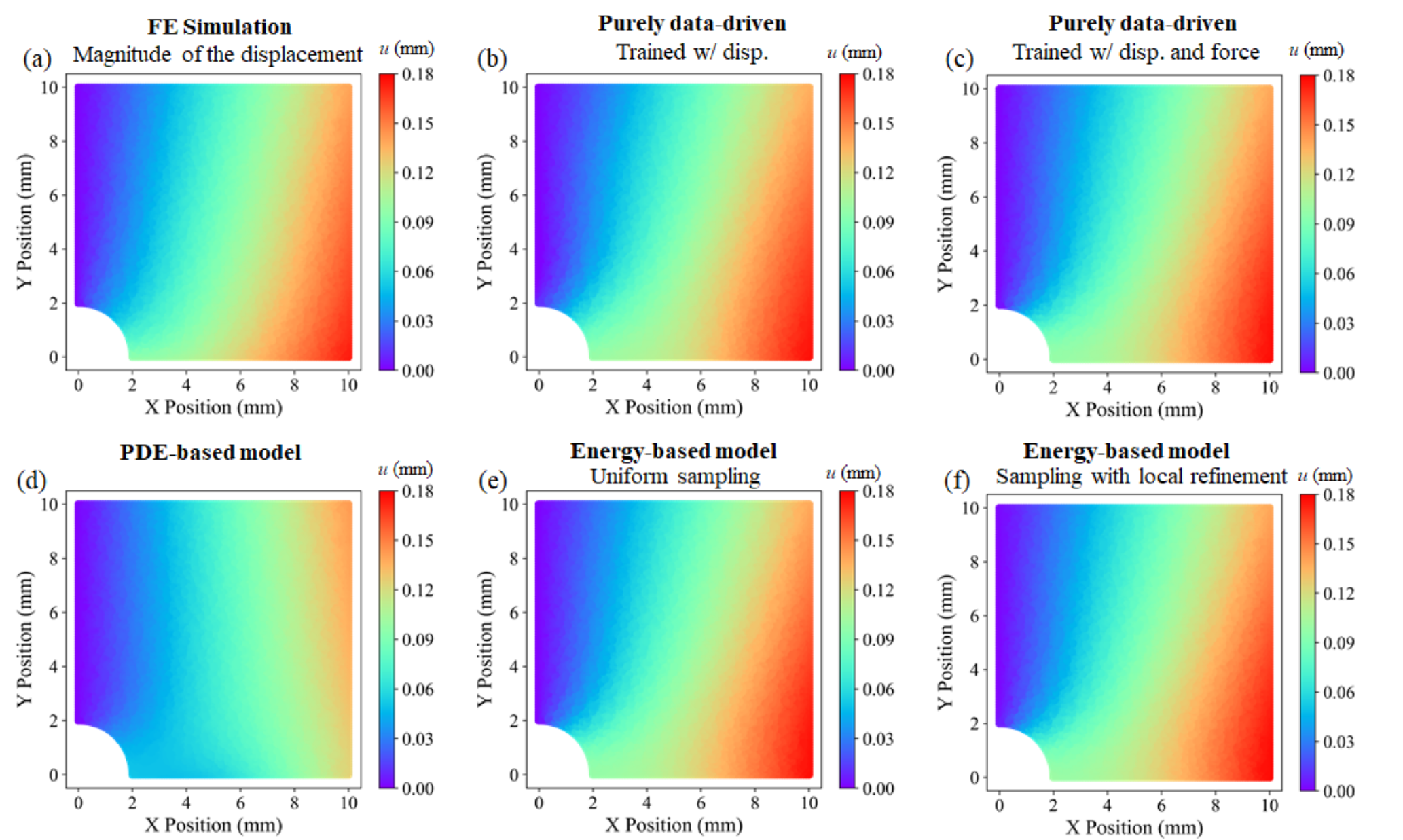}
        \caption{Horizontal displacement field prediction of  FE simulation (a), purely data-driven models with two different loss functions defined by: (b) mean square error of displacement field , (c) mean square error of displacement and membrane force fields, PDE-based model (d), and energy-based models trained with uniform sampling (e) and sampling with local refinement (f), for the central-hole tension. (For interpretation of the references to color in this figure legend, the reader is referred to the web version of this article.)}
        \label{fig:fig9}
    \end{figure}

\begin{figure}
        \centering
        \includegraphics[width=\columnwidth]{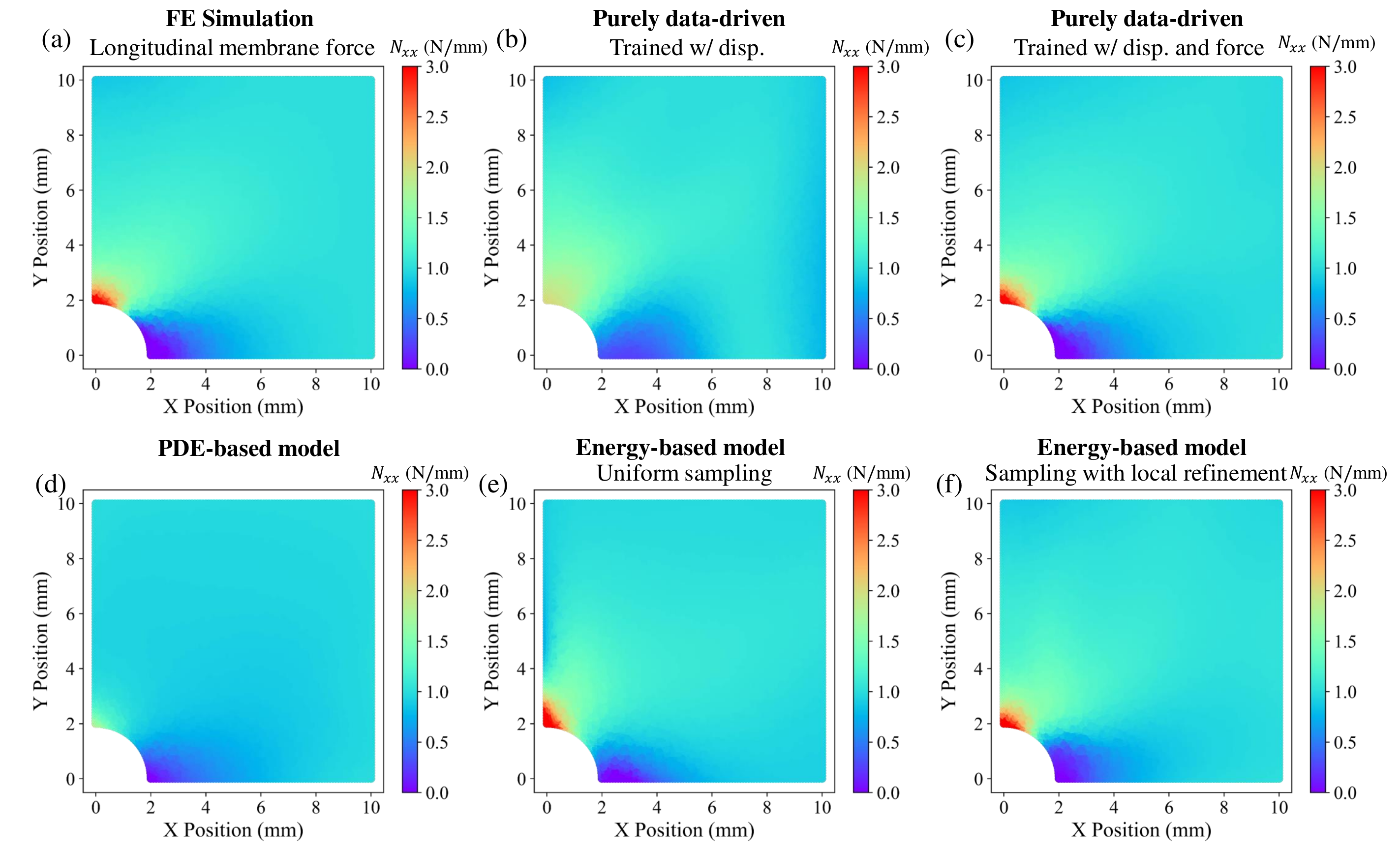}
        \caption{Longitudinal membrane force field prediction of  FE simulation (a), purely data-driven models with two different loss functions defined by: (b) mean square error of displacement field , (c) mean square error of displacement and membrane force fields, PDE-based model (d), and energy-based models trained with uniform sampling (e) and sampling with local refinement (f), for the central-hole tension. (For interpretation of the references to color in this figure legend, the reader is referred to the web version of this article.)}
        \label{fig:fig10}
    \end{figure}

    Similar to the previous case, we first performed a FE simulation as the reference and then trained a 5-hidden layer neural network (5 neurons each layer) with the displacement and membrane force fields from FE simulation results as purely data-driven cases. The neural network was trained in two ways: one only with the displacement field and the other with both the displacement field and the force field. A comparison between the predictions of the FE results and the two purely data-driven models is presented in Figure~{\ref{fig:fig9}}a, b, c about the displacement field and Figure~{\ref{fig:fig10}}a, b, c about the membrane force field. When the model is trained only with the displacement field, the error in the membrane force prediction is significant, especially around the central hole; when trained with both fields, the accuracy can be greatly improved. There are two important implications: 1) the central hole introduces stronger nonlinearity compared with the previous case, making the modeling more difficult; 2) a 5-layer network can well describe both fields, but the fitting will not be satisfactory unless the model is trained properly.
    
    We then trained the same 5-hidden layer network with the physics-guided loss functions, one based on PDEs and the other based on energy. The same training procedure as in the previous case was used. $\lambda_s=0.1$ was set for the PDE-based loss function. We found that both models can qualitatively predict the distribution of displacement field (see Figure~{\ref{fig:fig9}}d, e). Besides, the stress concentration phenomenon is also captured, but the concentration factor may not. The predicted distributions of the longitudinal membrane force are plotted in Figure~{\ref{fig:fig10}}d, e. The prediction by the energy-based model is the closest to the FE simulation result, significantly better than the PDE-based. It should be noted that, as demonstrated in the previous case, the PDE-based method can potentially achieve a similar accuracy as the energy-based method does by tuning the hyperparameters (training epochs, learning rates, sampling size, batch size weights in the loss function, etc). In practical, however, it is difficult to efficiently find the optimum hyperparameters. Attempts were made by increasing the sample size to an extremely large number 40,000, by tuning the weights in the loss function, by adjusting the batch size from 64 to 512, and by increasing the network to 20 layers (20 neurons each layer). However, none of these attempts returned satisfactory results. It is still an open question on how to optimize the hyperparameters for a strong nonlinear problem. Here, the unsatisfactory predictions are reported to show this weakness of the PDE-based approach.

\begin{figure}
        \centering
        \includegraphics[width=\columnwidth]{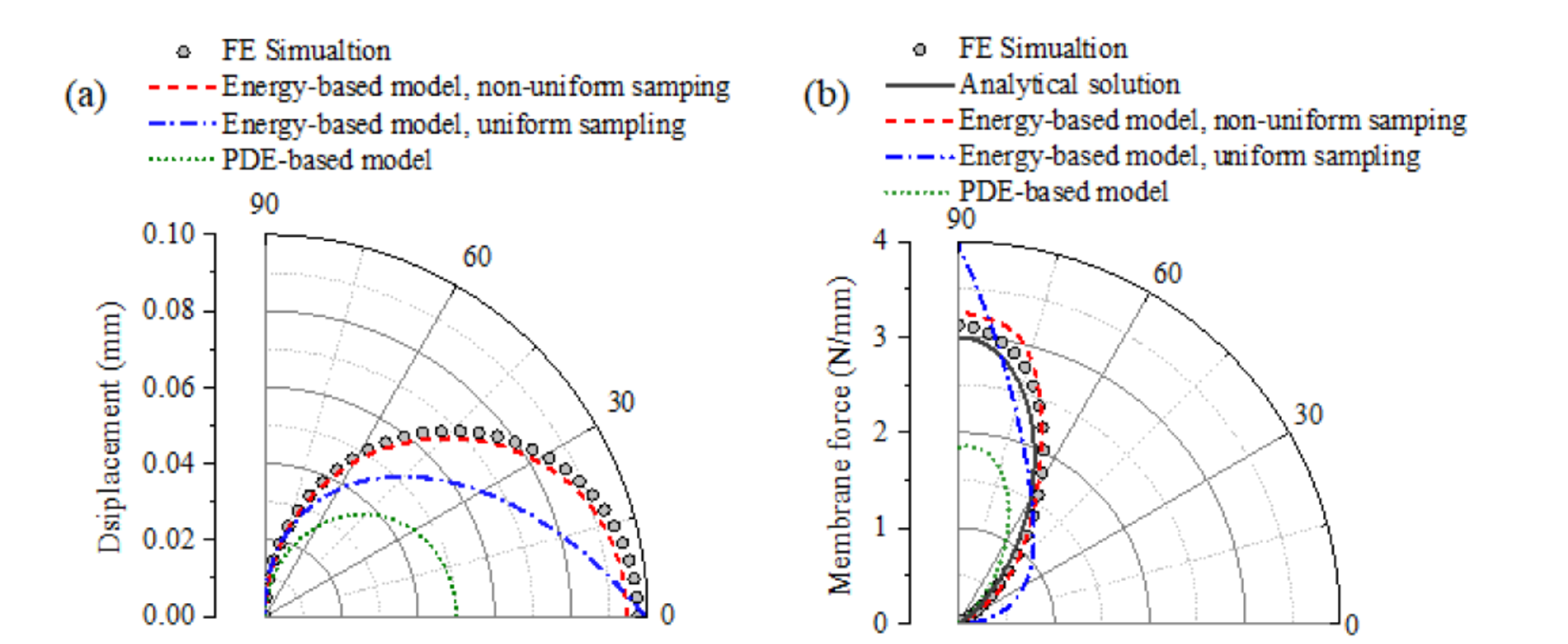}
        \caption{Predicted horizontal displacement (a) and longitudinal membrane force (b) of different models around the central hole.}
        \label{fig:fig11}
    \end{figure}

    A quantitative comparison between the physics-guided models and the FE results is performed by plotting the magnitude of $ N_{xx} $ along the edge of the central hole in the polar coordinate, as shown in Figure~{\ref{fig:fig11}}. It should be noted that a closed-form analytical solution exists for the uniaxial tension of an infinite large central-hole plate {\cite{sadd2009elasticity}}, which predicts $ N_{xx} = p(1 - 2 \cos 2 \phi){\sin\phi}^2 $. This analytical model is also plotted. At the same time, the $ R^2 $ values are summarized in Table~{\ref{tab:tab2}}. It is found that the error of the prediction around the stress concentration area is still notable for the energy-based model compared with the FE simulation and closed-form solution, although the predicted global distribution is qualitatively correct.
    
\begin{figure}
        \centering
        \includegraphics[width=\columnwidth]{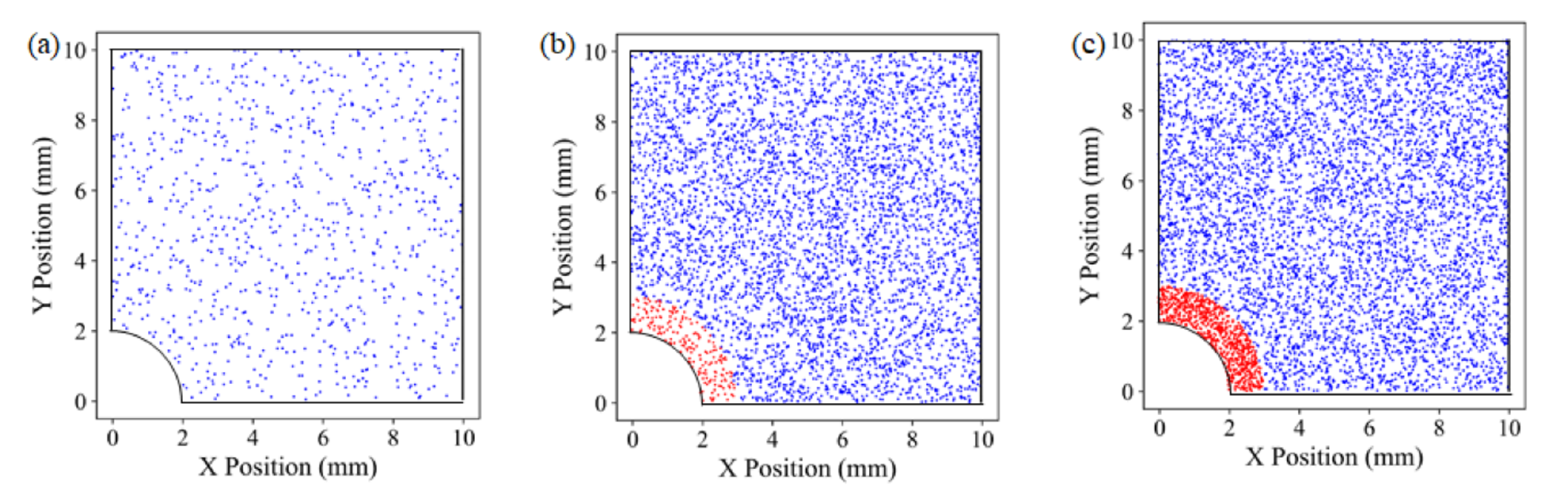}
        \caption{Illustration of different sampling size and strategy: (a) uniform sampling with small size and (b) large size, (c) sampling with local refinement (For interpretation of the references to color in this figure legend, the reader is referred to the web version of this article.)}
        \label{fig:CH sampling}
    \end{figure}

    To further improve the accuracy, we looked into the sampling strategy that has been revealed through the first in-plane tension example to be important to the numerical integration calculating the potential energy. First, it was found that the sampling size should be large enough to capture the edge of the central hole. As shown in Figure~{\ref{fig:CH sampling}}a and b, a small sample size cannot well capture the edge of central hole and 10,000 samples turned out to be sufficient for this example. Second, we proposed a sampling strategy with local refinement. Given that the stress concentration and high energy density area locates near the central hole, we divided the whole domain into two regions (indicated by red and blue in Figure~{\ref{fig:CH sampling}}c). The region around the central hole is sampled in higher density (local refinement). Within each region we performed the numerical integration, and the total potential energy is the sum of these two regions.
    
\begin{figure}
        \centering
        \includegraphics[width=0.9\columnwidth]{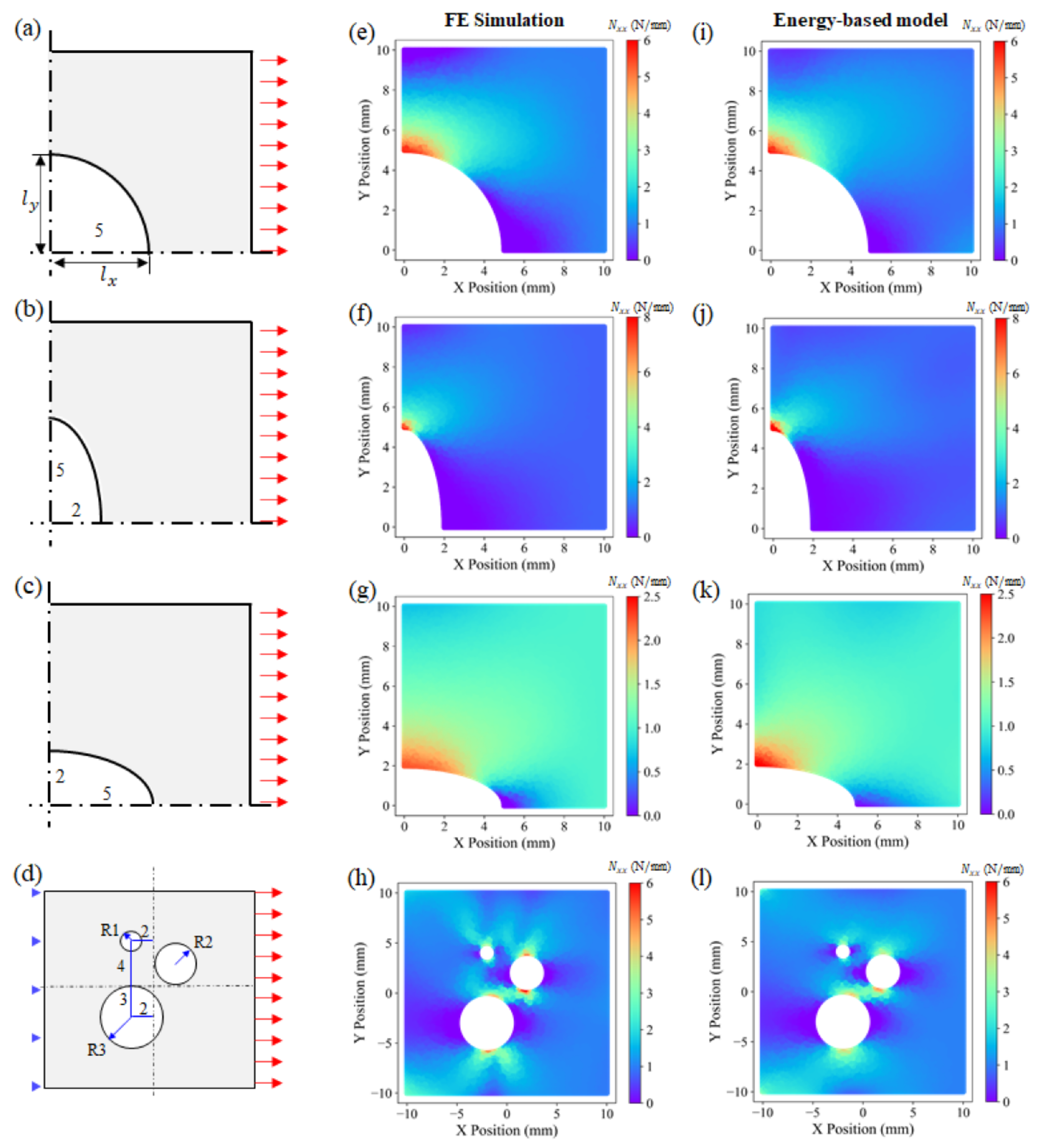}
        \caption{Comparison of FE simulation and energy-based model results for different shape and size central-hole tension: (a)  $l_{x}=l_{y}=4\  \text{mm}$, (b) $l_{x}=2\ \text{mm}, l_{y}=4\  \text{mm}$, (c) $l_{x}=4\ \text{mm}, l_{y}=2\  \text{mm}$, and (d) three holes. The membrane force fields predicted by the FE simulation of the four cases are respectively shown in (e), (f), (g), and (h). The predictions of the energy-based model are respectively shown in (i), (j), (k), and (l). (For interpretation of the references to color in this figure legend, the reader is referred to the web version of this article.)}
        \label{fig:fig12}
    \end{figure}

    It should also be pointed out that even the energy-based method cannot perfectly agree with the FE result. However, for such a nonlinear problem with stress concentration, we cannot fully trust the FE simulations as well. We expect a more persuasive comparison with the experimental data in future studies. Here, to further scrutinize the energy-based method, we applied it to three more cases with a central hole of different sizes and shapes. Meanwhile, a even more complicated case of three-hole plate was performed. All the results are shown in Figure~\ref{fig:fig12}. As expected, the stress concentration factor increases as the aspect ratio (\(l_{y}/l_{x}\)) of the central hole increases. The energy-based model can correctly capture this trend and is also able to give reasonable predictions of the full stress field. It is still seen that there is local deviations between the energy-based model and the FE simulation, but a prediction accuracy of over 95\% should have already met the requirement of most engineering applications.

\begin{table}[]
        \centering
        \caption{Coefficients of determination of the predicted displacement and membrane force fields for central-hole tension}
        \label{tab:tab2}
        \begin{tabular}{l|c|c|c|c|c|}
            \toprule
            Neural network     & \multicolumn{5}{c}{Coefficient of determination $(R^2)$} \\
                               & \(u_x\) & \(u_y\) & \(N_{xx}\) & \(N_{yy}\) & \(N_{xy}\) \\ \midrule
            Data-driven (w/ disp.) & 0.9996	& 0.9989	& 0.1274	& 0.4602	& 0.8028  \\
            Data-driven (w/ disp. and force) & 0.9990	& 0.9955	& 0.9946	& 0.9678	& 0.9807 \\
           PDE-based \((\lambda_s=0.1)\) & 0.6900	& 0.0559	& 0.2470	& -0.0163	& -0.1187 \\
             Energy-based (uniform sampling) & 0.9895	& 0.9883	& 0.9182	& 0.8672	& 0.9908\\ 
             Energy-based (w/ local refinement) & 0.9835	& 0.9754	& 0.8754	& 0.4770	& 0.8149\\
             \bottomrule
        \end{tabular}
    \end{table}

    \subsection{Out-of-plane deflection of a square plate}\label{out-of-plane-deflection-of-square-plate}

    The above two examples played as very strict comparisons of the purely data-driven approach and the two models with PDE-based and energy-based loss functions while using FE simulations as the reference. It has been demonstrated that both the PDE-based and the energy-based models can provide the high-accuracy predictions that are close to the FE results but the PDE-based heavility relies on the optimization of the hyperparameters. In the following two examples, we also compared the three ANN methods, and we reached the same conclusions. Therefore, the details of the comparison will not be presented for conciseness. Instead, we will focus on the energy-based method while still using FE simulation as the reference although it is not the exact solution.

\begin{figure}
        \centering
        \includegraphics[width=\columnwidth]{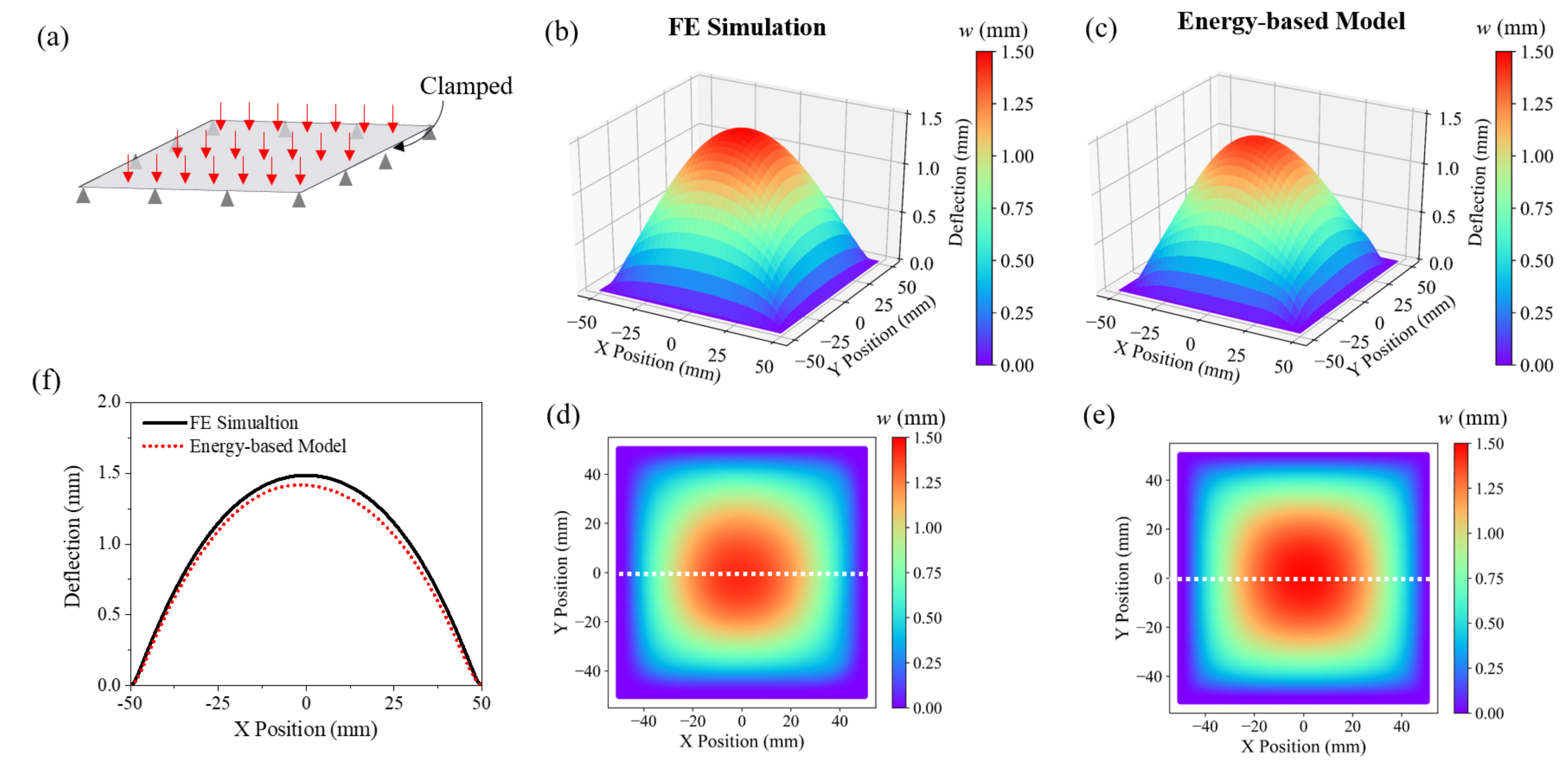}
        \caption{Loading case of plate deflection under out-of-plane pressure (a). The predicted 3D configuration and 2D out-of-plane displacement field after deflection are shown in (b) and (d) for FE simulation and (c) and (d) for energy-based model. The deflection of the central line is compared in (e) (For interpretation of the references to color in this figure legend, the reader is referred to the web version of this article.)}
        \label{fig:fig13}
    \end{figure}

    In this example, we consider the square plate deflection loading case under uniform out-of-plane pressure. Unlike the previous two, this case will involve both in-plane and out-of-plane deformation. As illustrated in Figure~\ref{fig:fig13}a, a 10 Pa uniform transverse distributed pressure is applied on a 100 mm \(\times\) 100 mm square plate whose four edges are clamped. The Young's modulus and Poisson's ratio of the plate is set as 70 MPa and 0.3, respectively. The governing PDEs are listed in Eq. \eqref{eq:eq9}. The boundary conditions are

    \begin{equation}
        \label{eq:3-9}
        u_x = 0,u_y = 0,w = 0,\frac{\partial w}{\partial n} = 0, \left( \text{at }x = \pm 50\ \text{or}\ y = \pm 50 \right).
    \end{equation}

    FE simulations with an element size of 0.1 mm are performed in Abaqus/standard with shell element (4-node doubly curved thin shell element, with reduced integration, hourglass control, and finite membrane strains). A 5-hidden layer neural network (5 neurons each layer) is trained with the energy-based loss function and the comparison of the predicted deflection field with the FE simulation is shown in Figure~\ref{fig:fig13}b-e. A quantitative comparison of the distribution along the central line is shown in Figure~\ref{fig:fig13}f. It is found that the energy-based algorithm can still provide a satisfactory prediction of the out-of-plane displacement. It should be noted that the small deviation between the energy-based method and the FE simulation cannot be fully ascribed to the computational error of the former. This is because our physics-guided neural network framework for elastic plates is constructed to implement the classic plate theory based on Kirchhoff hypotheses. In the FE simulations, the governing equations of the shell elements are slightly different, which depends on the integration algorithms and element size. For a stricter comparison, we can either employ experimental data or develop finite element simulations with the same classic plate theory.

    \subsection{Buckling of a square plate}\label{buckling-of-rectangular-plate} 

\begin{figure}
        \centering
        \includegraphics[width=\columnwidth]{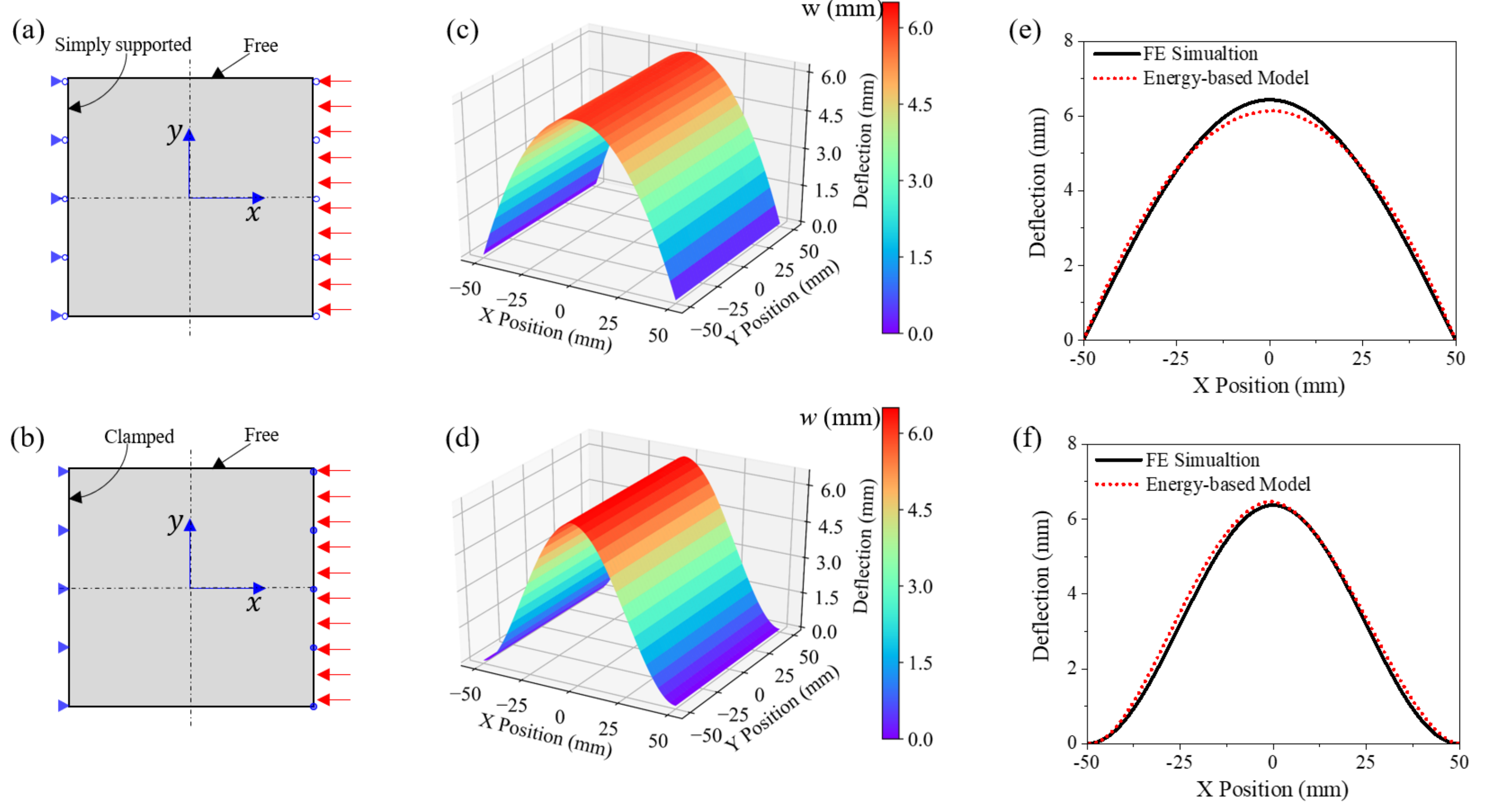}
        \caption{Prediction of the buckling of elastic plates with two different types of constraints: simply supported (a) and clamped (b). The buckled configurations are shown in (c) and (d). The deflections of the central line are compared with FE simulations (e) and (f). (For interpretation of the references to color in this figure legend, the reader is referred to the web version of this article.)}
        \label{fig:fig14}
    \end{figure}

    In the last example, we investigate the buckling of the same plate as the third example under in-plane compressive loads. Figure~\ref{fig:fig14}a and b show the two different boundary conditions that were studied. One has a simply-supported left edge (\(u_x = 0,w = 0,N_{yy}=0,M_{yy}=0\) at \(x = - 50\)), and the other clamped (\(u_x = 0,w = 0,N_{yy}=0,\partial w/\partial x = 0\) at \(x = - 50\)). In both cases, there is no out-of-plane load. Therefore, trivial solutions that only involve the in-plane deformation (i.e. no out-of-plane deflection, \(w = 0\)) exist because the trivial solutions always satisfy the out-of-plane governing equation. The deformation of the plate follows the trivial solutions when the load is sufficiently small, but as the load increases, there is a point \noindent where the plate will bifurcate into a more stable configuration (with lower potential energy) in a buckled shape. In plate theory, the first buckling mode is usually determined by seeking the lowest total potential energy. Therefore, we applied the energy-based model to predict it. The PDE-based loss function is not suitable for the buckling analysis since it inevitably converges to the trivial in-plane solution.

    For the neural network algorithms, a 5-hidden layer neural network (5 neurons each layer) with the energy-based loss function was constructed. The FE simulations were performed in Abaqus/standard to get the first buckling mode. Modal analysis was first conducted to obtain the different buckling modes. The first buckling mode configuration was then induced as the geometric imperfection with a maximum 0.01 mm transverse deviation and the model with the imperfection is used to simulate the in-plane compression with the implicit solver. The buckled configuration predicted by the neural network algorithms for the two cases are shown in Figure~\ref{fig:fig14}c and d, respectively. In addition, Figure~\ref{fig:fig14}e and f respectively compare the deflection of the central line with the FE simulations. We can see that the bulking configurations under two different boundary conditions are both correctly predicted.
    
\begin{figure}
        \centering
        \includegraphics[width=\columnwidth]{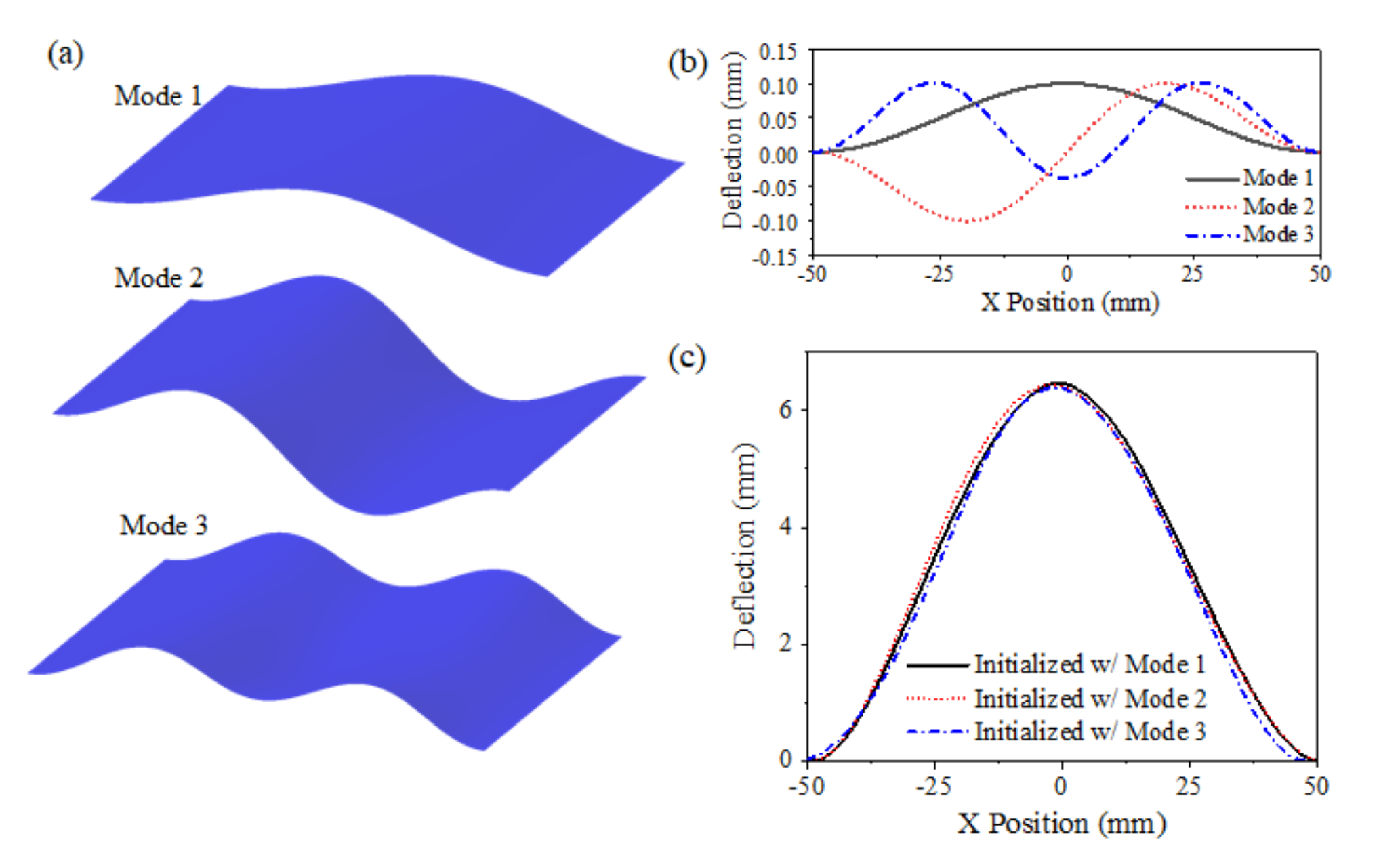}
        \caption{(a) Three different buckling modes under in-plane compression. Mode 1 corresponds to the lowest potential energy. (b) Deflection profiles of the three buckling modes for initializing the neural network. (c) Prediction of deflection when initialized with different deformation mode. The neural network is pre-trained to fit the initial deflection profiles.}
        \label{fig:buckle}
    \end{figure}

    The machine learning algorithms are designed to find the global minimum that corresponds to the first buckling mode in the studied case. It is intriguing whether the energy-based model can always converge to the global minimum or may find local minimums corresponding to higher buckling modes. We initialized the neural network to three different buckling modes: the first mode and two other higher modes (Figure~{\ref{fig:buckle}}a). This can be realized by pre-training the network to fit the initial configurations as shown in Figure~{\ref{fig:buckle}}b. Figure~{\ref{fig:buckle}}c presents the predicted deflections, where we can see that the final deformed profiles are almost the same regardless of the initial configuration. These results serve as a validation that the optimization algorithm that is being used in this study can find the global minimum. It also suggests that the energy-based approach is not able to find the higher modes.

\section{Discussions}\label{discussions}

\subsection{Comparison through the four loading conditions}\label{discussion_four_loading}

    In this study, we chose four typical loading conditions to validate the physics-guided neural network framework that we developed. Although they are all simple in terms of loading and boundary conditions, they together provided a rather comprehensive investigation of the accuracy of different approaches. The first example involved only in-plane deformation, but a nonlinear stretching force was applied. Among the four examples, this may be the simplest task for modeling. However, the differences between the data-driven approach and the physics-guided approaches and between the PDE-based and energy-based approaches were already clear. The second example was also in-plane, but geometrical nonlinearity was generated by introducing a circular hole at the center of the plate. Different aspect ratios of the central hole were investigated to obtain a wide range of the stress concentration factor, and a three-hole plate was solved to push the computational framework to its limit. We observed that there is a small difference between the FE simulation and energy-based model. However, it is promising to find that the accuracy of the energy-based method did not decrease as the stress concentration factor increased. In other words, this method is stable. The third example involved a pressure in the \(z\)-direction so that the out-of-plane governing equation could no longer be neglected. The energy-based method still provided a satisfactory prediction. The last example was more challenging due to instability. No out-of-plane load was applied, but out-of-plane deformation occurred through buckling. The energy-based method showed a great advantage over its PDE-based counterpart because the latter always converged to the trivial solution with \(w=0\). Therefore, these four examples covered almost all the important aspects of plate deformation. 

\subsection{Comparison between the PDE-based and energy-based approaches}\label{discussion_comparison_PDE_Energy}

    The major task of the present study is to compare the PDE-based and energy-based approaches of formulating the loss function. Although there is no absolute conclusion about which is better, it is clear that these two approaches have their own pros and cons. In this sub-section, we summarize them in three aspects: hyperparameters, sampling, and computational efficiency.
    
    \textbf{\textit{Hyperparameters}} -- The PDE-based loss function involves a larger number of hyperparameters than the energy-based does. This is a clear disadvantage of the PDE-based approach, although it was found through the first example of in-plane tension that the two approaches could achieve a similar accuracy as long as the hyperparameters could be determined properly. In the second in-plane example, we showed that it is difficult to find the optimum hyperparameters and that the prediction of the PDE-based approach could fail to match the results if the non-optimum hyperparameters are used.
    
    \textbf{\textit{Sampling}} -- The energy-based approach is significantly more dependent on the size and resolution of the samples used for training than the PDE-based, which is an important weakness. This is because the energy-based approach has to perform the numerical integration of the total potential energy, which relies on the discretion of the domain. In this sense, the energy-based approach is a ``mesh sensitive" tool although the concept of meshing is not explicitly stated. In the second in-plane example, this sampling strategy is close to a meshing step in conventional FE simulations. The difference is that there is not a strict requirement for the ``mesh" quality.
    
    \textbf{\textit{Computational efficiency}} -- To achieve the same accuracy that is sufficiently high to approximate the exact solution, the energy-based approach turns out to be more efficient. Here, the efficiency not only refers to the computation time that the algorithm takes to converge but also includes the time of the user to tune the hyperparameters.
    
\subsection{Fundamental differences between the PDE-based and energy-based loss functions}\label{discussion_difference}

    There are two prominent differences between the PDE-based and energy-based loss functions. The first lies in the order of the involved partial derivatives. The energy-based loss function deals with the strain components, which are functions of the first derivatives of the displacement field. By constructing the neural network to directly output the strain components, for example in the first example of in-plane tension, it is possible to avoid additional computational errors coming from the derivation process. On the contrary, the PDE-based loss function has to include the residuals in the PDEs and BCs simultaneously. Therefore, it is almost impossible to reduce the order of equations by computational treatments, and consequently, a number of derivation processes have to be performed in the algorithm, accumulating the computational error. The second difference between the PDE-based and energy-based loss functions is the number of the residual terms. The complete PDE-based loss function has to sum up a total number of eleven residuals coming from three equations and eight boundary conditions. However, the energy-based deals only one residual by introducing the penalty term into the total energy. Even if considering the penalty term as an independent quantity, we still have only two residuals for summing up. This is a big simplification. As a result of these two aspects, the PDE-based approach is less computationally efficient. 

    There are more fundamental underlying mechanisms behind these two computational differences. As already pointed out above, the energy approach and the PDE governing equations are mathematically equivalent. The energy-based loss (Eq. \eqref{eq:2-9}) and the PDE-based loss (Eq. \eqref{eq:2-5} are respectively formulated following these two approaches and, therefore, should also be equivalent. However, the equivalence could only be achieved when the weight ratios of the PDE-based loss, \(\lambda_{\text{s}}\) and \(\lambda_{\text{d}}\) , can be determined in advance to have the physical meanings of displacement and force, respectively. In other words, \(\lambda_{\text{s}}\)  and \(\mathcal{L}_{\text{Bcs}}\), \(\lambda_{\text{d}}\) and \(\mathcal{L}_{\text{Bcd}}\) should be two pairs of conjugate variables in terms of potential energy. However, this is impossible not only because such values of \(\lambda_{\text{s}}\) and \(\lambda_{\text{d}}\) are difficult to calculate but also because they are usually not uniform in all boundaries. In a practical application of neural network-based computational frameworks, \(\lambda_{\text{s}}\) and \(\lambda_{\text{d}}\) are usually chosen by the user with the help of a careful tuning procedure. Our study suggests that physics could guide the determination of the hyperparameters, particularly the weights of the residuals. This will be an important future topic which is worth a comprehensive investigation.

\subsection{Limitations of the proposed neural network framework}\label{discussion_limitations}

    We have seen that for loading cases with high nonlinearity there is still a relatively large deviation of the local predictions between the energy-based neural network model and the FE results. Although we have noted that FE results are not necessarily the exact solution and that it is unfair to attribute the deviation only to the computational error of the neural network-based algorithms, it is still necessary to point out the limitations of the proposed computational framework. 

    The first limitation is indicated by its name – the accuracy and the applicability of this physics-guided computational framework is largely determined by the physical laws that are implemented by human brains. In our study, the physical laws are the classic plate theory. It is based on the strong Kirchhoff hypotheses, which will lose the applicability for moderately thick plates. The framework was developed based on these hypotheses and, therefore, inherits its limitations.

    The second limitation stems from the fundamental and shadow neural network we have used. Its capability to approximate a highly non-uniform displacement or strain field is limited due to its simplicity. A deep neural network that involves a larger number of hidden layers is likely to increase accuracy. In the present study, we focused on the implementation of physics into machine learning algorithms and we did not try a deeper neural network due to the limitation in computational resources. Another approach to improve the approximation ability is to modify the structure of the neural network or seek for other machine learning models. For example, Wang et al. \cite{Wang2020mesh-free} added extra connections between non-adjacent layers to improve the approximation ability of the neural network.

\subsection{Special challenges for applications of neural network-based algorithms in predicting the mechanical responses of solids}\label{discussion_challenges}

    As mentioned in the literature survey in the introductory section, many initial successes of PGML or PINN algorithms have been achieved in modeling the dynamics of fluids as well as the mass and heat transfer. To model the mechanical responses of solids, a special challenge is that the variables of interest (stresses and strains) are highly tensorial. As a comparison, in many cases, only the pressure of a fluid is wanted. Consequently, more PDEs and BCs have to be implemented into the loss function, leading to a low accuracy of the PDE-based algorithms as mentioned in \ref{discussion_limitations}. This point becomes very clear through our present study – the governing equations are established in three directions, and each edge of the plate has four pairs of conjugates as BCs. It is also worth noting that our present study only considered elastic plates. To implement the plasticity theories, even the simplest model, there will be more intermediate state variables, thus creating more PDEs and ordinary differential equations (ODEs) to be solved. This will be one more special challenge for the neural network-based algorithms to take for predicting the deformation of solids.

\subsection{Future extensions of the proposed neural network framework}

    As a pilot study, we demonstrated that the energy-based neural network framework can provide a satisfactory prediction of the mechanical response of elastic plates that is as good as the FE simulation results. This seemingly easy conclusion may cause an underrating of the contribution of this work. The conclusion can be generalized – for a system that is governed by a large number of PDEs and BCs, if the principle of minimum potential energy is applicable, a machine learning algorithm designed to minimize the potential energy will be more effective and efficient than directly minimizing the total sum of all the residuals stemming from the PDEs and BCs. One potential extension of our neural network framework is modeling the thermodynamics of materials, which is also based on some energy indicators. While it is true that there is no evidence to prove the current accuracy of the framework is better than FE simulations, as the complexity of the system keeps increasing, we expect that the neural network framework will show more clear advantages over the FE simulations. One such potential application is the modeling of multiphysics and multiscale systems, lithium-ion batteries as a typical example. It is well-known that the FE simulations often suffer from a stringent criterion to get converged when dealing with these systems. Energy-based models are promising to relieve the convergence requirement to provide an approximate solution for practical applications. The mechanism behind it is a tradeoff between the modeling accuracy and the computational feasibility.

\section{Conclusion}

    In this study, we established a physics-guided neural network-based computational framework to predict the mechanical responses of elastic plates. The physical laws that were implemented into the algorithm were from the classic plate theory derived following the Kirchhoff hypotheses. The governing PDEs are the well-known FvK equations, which can be derived from the principle of virtual displacement. In our computational framework, a neural network was constructed to output the displacement fields (or strain fields in some cases) with the input of spatial coordinates. Three different ways of formulating the loss function were investigated. One was purely data-driven by comparing the predicted displacement field with the observed one from tests or FE simulations. The other two were based on the physical laws. The PDE-based loss function was the total sum of all the residuals stemming from the PDEs and BCs, and the energy-based simply used the total potential energy as its loss. The computational framework that we developed were then applied to four different types of loading conditions, including 1) the in-plane tension with non-uniformly distributed stretching force to study the effect of the nonlinearity from external loads, 2) the in-plane central-hole tension to investigate the nonlinearity from geometric imperfections, 3) the out-of-plane deflection to examine the capability of modeling the out-of-plane deformation, and 4) the buckling induced by uniaxial compression to validate the algorithm on instability analysis. In all the four cases, FE simulations with an extremely fine mesh size were performed as references. Through these validations and comparisons, the following conclusions can be drawn.
    
    \begin{enumerate}
        \item[1)] Both the PDE-based and the energy-based neural networks algorithms developed in this study can approximately predict the mechanical response of elastic plates with a satisfactory accuracy that is close to the FE simulations if the hyperparameters are properly tuned.
        \item[2)] The advantage of the model with the energy-based loss function is that it has a small number of hyperparameters and is computationally more efficient. Its disadvantage is that it relies on a large sampling size and a fine sampling resolution.
        \item[3)] The model with the PDE-based loss function has an advantage over the energy-based because it is less dependent on sampling and has the potential to be ``mesh-free".
        \item[4)] In order to achieve a good accuracy, the purely data-driven approach is suggested to be trained with data from both displacement and membrane force fields.
        \item[5)] The fundamental difference between the energy-based and PDE-based approaches largely stem from the determination of the weight in the loss function and the calculation of the derivatives. Deciphering the relationship between the weights and the physical meanings can potentially improve the PDE-based approach.
    \end{enumerate}

    It is optimistically expected that our energy-based neural network framework will have a wide spectrum of applications in future studies. Particularly, it provides an important energy-optimization inspiration for modeling complex engineering systems involving multiple scales and multiple physics.

\appendix
\renewcommand{\theequation}{A.\arabic{equation}}
\setcounter{equation}{0}
\section{Appendix}\label{appendix}
\subsection{Strain components}\label{Appendix A.1}

    Following the Kirchhoff hypotheses, the displacement field can be expressed as:

    \begin{equation}
        \label{eq:5-1}
            u_\alpha = u^0_\alpha - z \cdot w_{,\alpha},\ (\alpha=x,y).
    \end{equation}
    \noindent where $u_\alpha (\alpha=x,y)$ and $w$ denote the in-plane displacement and out-of-plane displacement, respectively. $u^0$ represents the corresponding displacement at the mid-plane (i.e. $u^0_\alpha(x,y,z)=u_\alpha(x,y,0)$).
    
    The general three-dimensional second-order nonlinear Green strains are

    \begin{equation}
        \label{eq:5-2}
        \varepsilon_{\alpha\beta} = \frac{1}{2}\left(u_{\alpha,\beta} + u_{\beta,\alpha} + u_{\gamma,\alpha}u_{\gamma,\beta} \right),\ (\alpha,\beta, \gamma =x,y,z).
    \end{equation}

    We consider the moderate deformation of a plate, meaning that the transverse (i.e. out-of-plane) displacement gradients \(u_{z,x} = w_{,x}\)and \(u_{z,y} = w_{,y}\) can be relatively large and the in-plane displacement gradients \(u_{\alpha,\beta}, (\alpha,\beta=x,y)\) are small due to the large width and length. The second-order terms in the Green strains can be therefore omitted except the \(w_{,\alpha\beta} (\alpha,\beta=x,y)\).

    Substitute Eq. \eqref{eq:5-1} into the above equation, the strains can subsequently be simplified to the following strains of the 2D plate theory,
    
    \begin{equation}
        \label{eq:5-4}
        \begin{aligned}
            \varepsilon_{\alpha\beta} &=  \frac{1}{2}\left(u_{\alpha,\beta}^0 + u_{\beta,\alpha}^0 + w_{,\alpha}w_{,\beta} -z \cdot w_{,\alpha\beta} \right), (\alpha,\beta=x,y), \\
            \varepsilon_{\gamma 3} & = 0, (\gamma=x,y,z).
        \end{aligned}
    \end{equation}

\subsection{Integration by parts}\label{Appendix A.2}

    The virtual strains are calculated from the virtual displacements according to Eq. \eqref{eq:eq2} and Eq. \eqref{eq:eq3}. For the first term in Eq. \eqref{eq:eq4} , applying integration by parts we have

    \begin{equation}
        \label{eq:5-5}
        \begin{aligned}
        \int_{\Omega}^{}{N_{\alpha\beta}\updelta\varepsilon_{\alpha\beta}^{0}}\text{d}x\text{d}y 
        & = \int_{\Omega}^{}{\frac{1}{2}N_{\alpha\beta} \left(\updelta u^0_{\alpha,\beta} + \updelta u^0_{\beta,\alpha} + \updelta w_{,\alpha}w_{,\beta} + w_{,\alpha} \updelta w_{,\beta} \right)} \\
        & = \int_{\Omega}^{}{N_{\alpha\beta} \left(\updelta u^0_{\alpha,\beta} + \updelta w_{,\alpha}w_{,\beta} \right)} \\
        & = \int_{\Gamma}^{}\left( N_{\alpha\beta}\updelta u^0_{\alpha}n_{\beta} + N_{\alpha\beta}w_{,\beta}\updelta w_{\gamma}n_{\alpha} \right) \text{d}s - \int_{\Omega}^{}{\left\lbrack N_{\alpha\beta,\beta}\updelta u^0_{\alpha} + \left( N_{\alpha\beta}w_{,\beta} \right)_{,\alpha}\updelta w \right\rbrack \text{d}x\text{d}y },
        \end{aligned}
    \end{equation}

    \noindent where the partial integration is applied to get the displacement variation instead of its gradient. \(\mathbf{n} = n_{x}\mathbf{e}_{x} + n_{y}\mathbf{e}_{y}\) is the outward normal on the boundary (\(n_{x}\) and \(n_{y}\) are the direction cosines of the unit normal). For the second term, we have

    \begin{equation}
        \label{eq:5-6}
        \begin{aligned}
        \int_{\Omega}^{}{M_{\alpha\beta}\updelta\kappa_{\alpha\beta}}\text{d}x\text{d}y&= - \int_{\Omega}^{}{M_{\alpha\beta}\updelta w_{,\alpha\beta}}\text{d}x\text{d}y\\
        & = - \int_{\Gamma}^{} M_{\alpha\beta}\updelta w_{,\alpha} n_{\beta} \text{d}s + \int_{\Omega}^{}{M_{\alpha\beta,\beta}\updelta w_{,\alpha}}\text{d}x\text{d}y \\ 
        &= - \int_{\Gamma}^{} M_{\alpha\beta}\updelta w_{,\alpha}  n_{\beta}\text{d}s + \int_{\Gamma}^{}{M_{\alpha\beta,\beta\ }\updelta w }n_{\alpha}\text{d}s - \int_{\Omega}^{}M_{\alpha\beta,\alpha\beta}\updelta w \text{d}x\text{d}y,
        \end{aligned}
    \end{equation}

    \noindent where the integration by parts is applied twice.

\subsection{Boundary conditions}\label{Appendix A.3}

    We perform a coordinate transformation between the global Cartesian \((x,y,z)\) coordinate and the local Cartesian coordinate \((n,s,r)\) (see Fig. 1),

    \begin{equation}
        \label{eq:5-8}
        \begin{bmatrix}
            x \\
            y \\
            z \\
            \end{bmatrix} = \begin{bmatrix}
            \cos\theta & - \sin\theta & 0 \\
            \sin\theta & \cos\theta & 0 \\
            0 & 0 & 1 \\
            \end{bmatrix}\begin{bmatrix}
            n \\
            s \\
            r \\
            \end{bmatrix} = \begin{bmatrix}
            n_{x} & -n_{y} & 0 \\
            n_{y} & n_{x} & 0 \\
            0 & 0 & 1 \\
            \end{bmatrix}\begin{bmatrix}
            n \\
            s \\
            r \\
            \end{bmatrix},
    \end{equation}

    \noindent where \(\theta\) is the angle between the global \(x\) axis and the local \(n\) axis along the counterclockwise direction. The displacements and stresses under the two coordinates are related by

    \begin{equation}
        \label{eq:5-9}
        \begin{bmatrix}
            u_{0} \\
            v_{0} \\
            w \\
            \end{bmatrix} = \begin{bmatrix}
            \cos\theta & - \sin\theta & 0 \\
            \sin\theta & \cos\theta & 0 \\
            0 & 0 & 1 \\
            \end{bmatrix}\begin{bmatrix}
            u_{0n} \\
            u_{0s} \\
            w_{0r} \\
            \end{bmatrix},
    \end{equation}

    \begin{equation}
        \label{eq:5-10}
        \begin{bmatrix}
            \sigma_{\text{nn}} \\
            \sigma_{\text{ns}} \\
        \end{bmatrix} = \begin{bmatrix}
            n_{x}^{2} & n_{y}^{2} & 2n_{x}n_{y} \\
             - n_{x}n_{y} & n_{x}n_{y} & n_{x}^{2} - n_{y}^{2} \\
        \end{bmatrix}\begin{bmatrix}
            \sigma_{xx} \\
            \sigma_{yy} \\
            \sigma_{xy} \\
        \end{bmatrix}.
    \end{equation}

    According to the above relations, the stress boundary integrands in Eq. \eqref{eq:eq8} can be rewritten with quantities under the local coordinate,

    \begin{equation}
        \label{eq:5-11}
        \begin{aligned}
        ( N_{xx}n_{x} &+ N_{xy}n_{y}) \updelta u_{0}  + \left( N_{yy}n_{y} + N_{xy}n_{x} \right)\updelta v_{0} \\
        &= \left( N_{xx}n_{x} + N_{xy}n_{y} \right)\left( n_{x}\updelta u_{0n} - n_{y}\updelta u_{0s} \right) 
         + \left( N_{yy}n_{y} + N_{xy}n_{x} \right)\left( n_{y}\updelta u_{0n} + n_{x}\updelta u_{0s} \right) \\
        & = \left( N_{xx}n_{x}^{2} + 2N_{xy}n_{x}n_{y} + N_{yy}n_{y}^{2} \right)\updelta u_{0n} 
         + \left\lbrack N_{yy}n_{x}n_{y}{- N}_{xx}n_{x}n_{y} + N_{xy}\left( n_{x}^{2} - n_{y}^{2} \right) \right\rbrack\updelta u_{0s} \\
        & = N_{\text{nn}}\updelta u_{0n} + N_{\text{ns}}\updelta u_{0s},
        \end{aligned}
    \end{equation}

    and

    \begin{equation}
        \label{eq:5-12}
        - \left( M_{yy}n_{y} + M_{xx}n_{x} \right)\frac{\partial\updelta w}{\partial y} - \left( M_{xx}n_{y} + M_{xx}n_{x} \right)\frac{\partial \updelta w}{\partial x} = - M_{nn}\frac{\partial \updelta w}{\partial n} - M_{ns}\frac{\partial \updelta w}{\partial s}.
    \end{equation}

    To fix the inconsistency, integration by parts is applied,

    \begin{equation}
        \label{eq:5-13}
        \int_{\Gamma_{\sigma}}^{}{M_{ns}\frac{\partial \updelta w}{\partial s}\text{d}s} = - \left\lbrack M_{ns}\updelta w \right\rbrack_{\Gamma_{\sigma}} + \int_{\Gamma_{\sigma}}^{}{\frac{\partial M_{ns}}{\partial s}\updelta w\text{d}s} . 
    \end{equation}

    \(\left\lbrack M_{ns}\updelta w \right\rbrack_{\Gamma_{\sigma}}\)is zero when the stress boundary is closed or \(M_{ns} = 0\). Then we can get the Eq. \eqref{eq:eq12}.

    \section*{Acknowledgment} 
    J.Z. and W.L. are grateful to the support by AVL, Hyundai, Murata, Tesla, Toyota North America, Volkswagen/Audi/Porsche, and other industrial partners through the MIT Industrial Battery Consortium. M.Z.B is grateful to the support by Toyota Research Institute through the D3BATT Center on Data-Driven-Design of Rechargeable Batteries. Thanks are also due to the MIT-Indonesia Seed Fund to support J.Z.'s postdoctoral study.

\printbibliography

@article{Egmont-Petersen2002Image,
	journal={Pattern Recognition},
	doi={10.1016/S0031-3203(01)00178-9},
	issn=00313203,
	title={Image processing with neural networks- A review},
	author={Egmont-Petersen, Michael and de Ridder, Dick and Handels, Heinz},
	date=2002,
	year=2002,
}

@article{Rawat2017Deep,
	journal={Neural Computation},
	doi={10.1162/NECO_a_00990},
	issn={1530888X},
	pmid=28599112,
	title={Deep convolutional neural networks for image classification: A comprehensive review},
	author={Rawat, Waseem and Wang, Zenghui},
	date=2017,
	year=2017,
}

@article{French2002Introduction,
	journal={Biological Psychology},
	doi={10.1016/s0301-0511(02)00012-1},
	issn=03010511,
	title={Introduction to Neural and Cognitive Modeling},
	author={French, Robert M},
	date=2002,
	year=2002,
}

@article{Libbrecht2015Machine,
	journal={Nature Reviews Genetics},
	doi={10.1038/nrg3920},
	issn=14710064,
	pmid=25948244,
	title={Machine learning applications in genetics and genomics},
	author={Libbrecht, Maxwell W. and Noble, William Stafford},
	date=2015,
	year=2015,
}

@article{Lo2018Machine,
	journal={Drug Discovery Today},
	doi={10.1016/j.drudis.2018.05.010},
	issn=18785832,
	pmid=29750902,
	title={Machine learning in chemoinformatics and drug discovery},
	author={Lo, Yu Chen and Rensi, Stefano E. and Torng, Wen and Altman, Russ B.},
	date=2018,
	year=2018,
}

@article{Ramprasad2017Machine,
	journal={npj Computational Materials},
	doi={10.1038/s41524-017-0056-5},
	issn=20573960,
	title={Machine learning in materials informatics: Recent applications and prospects},
	author={Ramprasad, Rampi and Batra, Rohit and Pilania, Ghanshyam and Mannodi-Kanakkithodi, Arun and Kim, Chiho},
	date=2017,
	year=2017,
}

@article{Alber2019Integrating,
	journal={npj Digital Medicine},
	doi={10.1038/s41746-019-0193-y},
	issn={2398-6352},
	title={Integrating machine learning and multiscale modeling—perspectives, challenges, and opportunities in the biological, biomedical, and behavioral sciences},
	author={Alber, Mark and Buganza Tepole, Adrian and Cannon, William R. and De, Suvranu and Dura-Bernal, Salvador and Garikipati, Krishna and Karniadakis, George and Lytton, William W. and Perdikaris, Paris and Petzold, Linda and Kuhl, Ellen},
	date=2019,
	year=2019,
}

@article{Han2018Solving,
	journal={Proceedings of the National Academy of Sciences of the United States of America},
	doi={10.1073/pnas.1718942115},
	issn=10916490,
	number=34,
	pmid=30082389,
	title={Solving high-dimensional partial differential equations using deep learning},
	volume=115,
	author={Han, Jiequn and Jentzen, Arnulf and Weinan, E.},
	pages={8505--8510},
	date=2018,
	year=2018,
}

@article{Severson2019Data-driven,
	journal={Nature Energy},
	doi={10.1038/s41560-019-0356-8},
	isbn=4156001903,
	issn=20587546,
	number=5,
	publisher={Springer US},
	title={Data-driven prediction of battery cycle life before capacity degradation},
	volume=4,
	author={Severson, Kristen A. and Attia, Peter M. and Jin, Norman and Perkins, Nicholas and Jiang, Benben and Yang, Zi and Chen, Michael H. and Aykol, Muratahan and Herring, Patrick K. and Fraggedakis, Dimitrios and Bazant, Martin Z. and Harris, Stephen J. and Chueh, William C. and Braatz, Richard D.},
	pages={383--391},
	date=2019,
	year=2019,
}

@article{Famili1997Data,
	journal={Intelligent Data Analysis},
	doi={10.3233/IDA-1997-1102},
	issn=15714128,
	title={Data preprocessing and intelligent data analysis},
	author={Famili, A. and Shen, Wei Min and Weber, Richard and Simoudis, Evangelos},
	date=1997,
	year=1997,
}

@article{Sirignano2018DGM:,
	journal={Journal of Computational Physics},
	doi={10.1016/j.jcp.2018.08.029},
	issn=10902716,
	number={Dms 1550918},
	title={DGM: A deep learning algorithm for solving partial differential equations},
	volume=375,
	author={Sirignano, Justin and Spiliopoulos, Konstantinos},
	pages={1339--1364},
	date=2018,
	year=2018,
}

@article{Wang2020mesh-free,
	journal={Journal of Computational Physics},
	doi={10.1016/j.jcp.2019.108963},
	issn=10902716,
	title={A mesh-free method for interface problems using the deep learning approach},
	volume=400,
	author={Wang, Zhongjian and Zhang, Zhiwen},
	date=2020,
	year=2020,
}

@article{Weinan2018Deep,
	journal={Communications in Mathematics and Statistics},
	doi={10.1007/s40304-018-0127-z},
	issn={2194671X},
	number=1,
	title={The Deep Ritz Method: A Deep Learning-Based Numerical Algorithm for Solving Variational Problems},
	volume=6,
	author={Weinan, E. and Yu, Bing},
	pages={1--14},
	date=2018,
	year=2018,
}

@article{Raissi2019Physics-informed,
	journal={Journal of Computational Physics},
	doi={10.1016/j.jcp.2018.10.045},
	issn=10902716,
	publisher={Elsevier Inc.},
	title={Physics-informed neural networks: A deep learning framework for solving forward and inverse problems involving nonlinear partial differential equations},
	volume=378,
	author={Raissi, M. and Perdikaris, P. and Karniadakis, G. E.},
	pages={686--707},
	date=2019,
	year=2019,
}

@article{Wang2017Physics-informed,
	journal={Physical Review Fluids},
	doi={10.1103/PhysRevFluids.2.034603},
	issn={2469990X},
	number=3,
	title={Physics-informed machine learning approach for reconstructing Reynolds stress modeling discrepancies based on DNS data},
	volume=2,
	author={Wang, Jian Xun and Wu, Jin Long and Xiao, Heng},
	pages={1--22},
	date=2017,
	year=2017,
}

@article{Samaniego2020energy,
	journal={Computer Methods in Applied Mechanics and Engineering},
	doi={10.1016/j.cma.2019.112790},
	issn=00457825,
	publisher={Elsevier B.V.},
	title={An energy approach to the solution of partial differential equations in computational mechanics via machine learning: Concepts, implementation and applications},
	volume=362,
	author={Samaniego, E. and Anitescu, C. and Goswami, S. and Nguyen-Thanh, V. M. and Guo, H. and Hamdia, K. and Zhuang, X. and Rabczuk, T.},
	pages=112790,
	date=2020,
	year=2020,
}

@article{Karpatne2017Theory-guided,
	journal={IEEE Transactions on Knowledge and Data Engineering},
	doi={10.1109/TKDE.2017.2720168},
	issn=10414347,
	number=10,
	title={Theory-guided data science: A new paradigm for scientific discovery from data},
	volume=29,
	author={Karpatne, Anuj and Atluri, Gowtham and Faghmous, James H. and Steinbach, Michael and Banerjee, Arindam and Ganguly, Auroop and Shekhar, Shashi and Samatova, Nagiza and Kumar, Vipin},
	pages={2318--2331},
	date=2017,
	year=2017,
}

@article{Li2019Data-Driven,
	journal={Joule},
	doi={10.1016/j.joule.2019.07.026},
	issn=25424351,
	publisher={Elsevier Inc.},
	title={Data-Driven Safety Envelope of Lithium-Ion Batteries for Electric Vehicles},
	author={Li, Wei and Zhu, Juner and Xia, Yong and Gorji, Maysam B. and Wierzbicki, Tomasz},
	pages={1--13},
	date=2019,
	year=2019,
}

@misc{Chen2020Direct,
	title={Direct prediction of phonon density of states with Euclidean neural network},
	author={Chen, Zhantao and Andrejevic, Nina and Smidt, Tess and Ding, Zhiwei and Chi, Yen-Ting and Nguyen, Quynh T. and Alatas, Ahmet and Kong, Jing and Li, Mingda},
	pages={1--18},
	date=2020,
	year=2020,
    Eprint={arXiv:2009.05163},
}

@article{Zhang2018DeePCG:,
	journal={Journal of Chemical Physics},
	doi={10.1063/1.5027645},
	issn=00219606,
	pmid=30037247,
	title={DeePCG: Constructing coarse-grained models via deep neural networks},
	author={Zhang, Linfeng and Han, Jiequn and Wang, Han and Car, Roberto and Weinan, Weinan E.},
	date=2018,
	year=2018,
}

@article{Darbon2020Overcoming,
	journal={Research in Mathematical Sciences},
	doi={10.1007/s40687-020-00215-6},
	issn=21979847,
	title={Overcoming the curse of dimensionality for some Hamilton–Jacobi partial differential equations via neural network architectures},
	author={Darbon, Jérôme and Langlois, Gabriel P. and Meng, Tingwei},
	date=2020,
	year=2020,
}

@article{Lu2019DeepXDE:,
	title={DeepXDE: A deep learning library for solving differential equations},
	author={Lu, Lu and Meng, Xuhui and Mao, Zhiping and Karniadakis, George E.},
	pages={1--21},
	date=2019,
	year=2019,
}

@article{Zhao2020Learning,
	journal={Physical Review Letters},
	doi={10.1103/PhysRevLett.124.060201},
	issn=10797114,
	pmid=32109085,
	title={Learning the Physics of Pattern Formation from Images},
	author={Zhao, Hongbo and Storey, Brian D. and Braatz, Richard D. and Bazant, Martin Z.},
	date=2020,
	year=2020,
}

@article{Effendy2020Analysis,
	journal={Journal of The Electrochemical Society},
	doi={10.1149/1945-7111/ab9c82},
	issn={1945-7111},
	title={Analysis, Design, and Generalization of Electrochemical Impedance Spectroscopy (EIS) Inversion Algorithms},
	author={Effendy, Surya and Song, Juhyun and Bazant, Martin Z.},
	date=2020,
	year=2020,
}

@article{E2020Integrating,
	title={Integrating Machine Learning with Physics-Based Modeling},
	author={E, Weinan and Han, Jiequn and Zhang, Linfeng},
	pages={1--23},
	date=2020,
	year=2020,
}

@article{Qian2020Lift,
	journal={Physica D: Nonlinear Phenomena},
	doi={10.1016/j.physd.2020.132401},
	issn=01672789,
	title={Lift \& Learn: Physics-informed machine learning for large-scale nonlinear dynamical systems},
	author={Qian, Elizabeth and Kramer, Boris and Peherstorfer, Benjamin and Willcox, Karen},
	date=2020,
	year=2020,
}

@article{Haghighat2020deep,
	title={A deep learning framework for solution and discovery in solid mechanics},
	author={Haghighat, Ehsan and Raissi, Maziar and Moure, Adrian and Gomez, Hector and Juanes, Ruben},
	date=2020,
	year=2020,
}

@article{Wu2020recurrent,
	journal={Computer Methods in Applied Mechanics and Engineering},
	doi={10.1016/j.cma.2020.113234},
	issn=00457825,
	title={A recurrent neural network-accelerated multi-scale model for elasto-plastic heterogeneous materials subjected to random cyclic and non-proportional loading paths},
	author={Wu, Ling and Nguyen, Van Dung and Kilingar, Nanda Gopala and Noels, Ludovic},
	date=2020,
	year=2020,
}

@article{Huang2020machine,
	journal={Computer Methods in Applied Mechanics and Engineering},
	doi={10.1016/j.cma.2020.113008},
	issn=00457825,
	publisher={Elsevier B.V.},
	title={A machine learning based plasticity model using proper orthogonal decomposition},
	volume=365,
	author={Huang, Dengpeng and Fuhg, Jan Niklas and Weißenfels, Christian and Wriggers, Peter},
	pages=113008,
	date=2020,
	year=2020,
}

@book{Foppl1899Vorlesungen,
	publisher={BG Teubner},
	title={Vorlesungen über technische Mechanik},
	volume=4,
	author={Föppl, August},
	date=1899,
	year=1899,
}

@incollection{Karman1907Festigkeitsprobleme,
	booktitle={Mechanik},
	doi={10.1007/978-3-663-16028-1_5},
	title={Festigkeitsprobleme im Maschinenbau},
	author={K{\'a}rm{\'a}n, Th V},
    pages={311--385},
    year={1907},
    publisher={Springer}
}

@article{Zhu2018Stretch-induced,
	journal={International Journal of Solids and Structures},
	doi={10.1016/j.ijsolstr.2018.02.005},
	issn=00207683,
	publisher={Elsevier Ltd},
	title={Stretch-induced wrinkling of highly orthotropic thin films},
	volume={139-140},
	author={Zhu, Juner and Zhang, Xiaowei and Wierzbicki, Tomasz},
	pages={238--249},
	date=2018,
	year=2018,
}

@article{Cerda2002Thin,
	journal={Nature},
	doi={10.1038/419579b},
	issn=00280836,
	pmid=12374968,
	title={Thin films: Wrinkling of an elastic sheet under tension},
	author={Cerda, E. and Ravi-Chandar, K. and Mahadevan, L.},
	date=2002,
	year=2002,
}

@article{Puntel2011Wrinkling,
	journal={Journal of Elasticity},
	doi={10.1007/s10659-010-9290-5},
	issn=03743535,
	title={Wrinkling of a stretched thin sheet},
	author={Puntel, Eric and Deseri, Luca and Fried, Eliot},
	date=2011,
	year=2011,
}

@article{Nayyar2011Stretch-induced,
	journal={International Journal of Solids and Structures},
	doi={10.1016/j.ijsolstr.2011.09.004},
	issn=00207683,
	title={Stretch-induced stress patterns and wrinkles in hyperelastic thin sheets},
	author={Nayyar, Vishal and Ravi-Chandar, K. and Huang, Rui},
	date=2011,
	year=2011,
}

@article{Sipos2016Disappearance,
	journal={International Journal of Solids and Structures},
	doi={10.1016/j.ijsolstr.2016.07.021},
	issn=00207683,
	title={Disappearance of stretch-induced wrinkles of thin sheets: A study of orthotropic films},
	author={Sipos, András A. and Fehér, Eszter},
	date=2016,
	year=2016,
}

@book{Reddy2006,
author = {Reddy, Junuthula Narasimha},
booktitle = {Theory and Analysis of Elastic Plates and Shells},
doi = {10.1201/9780849384165},
title = {{Theory and Analysis of Elastic Plates and Shells}},
year = {2006},
publisher={CRC press}
}

@misc{abadi2016tensorflow,
  title={Tensorflow: A system for large-scale machine learning},
  author={Abadi, Mart{\'\i}n and Barham, Paul and Chen, Jianmin and Chen, Zhifeng and Davis, Andy and Dean, Jeffrey and Devin, Matthieu and Ghemawat, Sanjay and Irving, Geoffrey and Isard, Michael and others},
  booktitle={12th $\{$USENIX$\}$ symposium on operating systems design and implementation ($\{$OSDI$\}$ 16)},
  pages={265--283},
  year={2016}
}

@misc{paszke2017automatic,
  title={Automatic differentiation in pytorch},
  author={Paszke, Adam and Gross, Sam and Chintala, Soumith and Chanan, Gregory and Yang, Edward and DeVito, Zachary and Lin, Zeming and Desmaison, Alban and Antiga, Luca and Lerer, Adam},
  year={2017}
}

@article{cybenko1989approximation,
  title={Approximation by superpositions of a sigmoidal function},
  author={Cybenko, George},
  journal={Mathematics of control, signals and systems},
  volume={2},
  number={4},
  pages={303--314},
  year={1989},
  publisher={Springer}
}

@book{sadd2009elasticity,
  title={Elasticity: theory, applications, and numerics},
  author={Sadd, Martin H},
  year={2009},
  publisher={Academic Press}
}

@article{long2019pde,
  title={PDE-Net 2.0: Learning PDEs from data with a numeric-symbolic hybrid deep network},
  author={Long, Zichao and Lu, Yiping and Dong, Bin},
  journal={Journal of Computational Physics},
  volume={399},
  pages={108925},
  year={2019},
  publisher={Elsevier}
}
\end{document}